\documentclass[aip,jcp,preprint]{revtex4-1}
\usepackage{graphicx}
\usepackage{rotating}
\usepackage{amsfonts,amsmath,latexsym,bbm,dsfont}
\usepackage{chemformula}
\usepackage[hidelinks]{hyperref}

\usepackage{color}

\newcommand{\nl}{\nonumber \\ & }

\newcommand{\beq}{\begin{eqnarray}}
\newcommand{\eeq}{\end{eqnarray}}
\newcommand{\half}{\frac{1}{2}}

\newcommand{\op}{\hat}

\newcommand{\dg}{\dagger}
\newcommand{\lk}{\left}
\newcommand{\rk}{\right}

\newcommand{\chg}[1]{{#1}}

\newcommand{\molpro}{\textsc{Molpro} }
\def\RCS$#1: #2 ${\@namedef{RCS#1}{#2}\typeout{RCS #1: #2}}
\mathchardef\lt="313C \mathchardef\gt="313E
\mathcode`<="4268 \mathcode`>="5269

\renewcommand\vec\mathbf

\begin{document}

\author{Thomas Schraivogel}
\email{t.schraivogel@fkf.mpg.de}
\affiliation{Max Planck Institute for Solid State Research, Heisenbergstra\ss e 1, 70569 Stuttgart, Germany}
\author{Evelin Martine Christlmaier}
\affiliation{Max Planck Institute for Solid State Research, Heisenbergstra\ss e 1, 70569 Stuttgart, Germany}
\author{Pablo L\'opez R\'ios}
\affiliation{Max Planck Institute for Solid State Research, Heisenbergstra\ss e 1, 70569 Stuttgart, Germany}
\author{Ali Alavi}
\affiliation{Max Planck Institute for Solid State Research, Heisenbergstra\ss e 1, 70569 Stuttgart, Germany}
\affiliation{Yusuf Hamied Department of Chemistry, University of Cambridge, Lensfield Road, Cambridge CB2 1EW, United Kingdom}
\author{Daniel Kats}
\email{d.kats@fkf.mpg.de}
\affiliation{Max Planck Institute for Solid State Research, Heisenbergstra\ss e 1, 70569 Stuttgart, Germany}

\title{Transcorrelated coupled cluster methods. II. Molecular systems}
\begin{abstract}
We demonstrate the accuracy of ground-state energies of the transcorrelated Hamiltonian, employing sophisticated Jastrow factors obtained from variational Monte Carlo, together with the coupled cluster and distinguishable cluster methods at the level of singles and doubles excitations.
Our results show that already with the cc-pVTZ basis the transcorrelated distinguishable cluster method gets close to 
complete basis limit and near full configuration interaction quality values for relative energies of over thirty atoms and molecules.
To gauge the performance in different correlation regimes we also investigate the breaking of the nitrogen molecule 
with transcorrelated coupled cluster methods.
Numerical evidence is presented to further justify an efficient way to incorporate the major effects coming from the three-body integrals without 
explicitly introducing them into the amplitude equations.
\end{abstract}

\maketitle 
\section{Introduction}
The coupled cluster (CC) theory\cite{cizek66,purvis82} provides a size-extensive, systematically improvable 
hierarchy of methods to approach the full configuration interaction~(FCI) limit.\cite{bartlett07}
Highly accurate ab-initio calculations with errors below 1 kJ/mol ("subchemical" accuracy) 
often require coupled cluster with singles, doubles, triples and quadruples\cite{kucharski92} (CCSDTQ) 
or perturbative quadruples\cite{bomble05} CCSDT(Q) and large basis sets,
which make the calculations prohibitively expensive for all but the smallest chemical moieties.\cite{tajti04} \newline
The root cause for the need of large basis sets is that many one-electron Gaussian type functions are needed to describe the Coulomb holes in the wave function accurately.
Coulomb holes occur in regions of the wave function where two electrons come close to each other and are caused by the requirement that the
wave function satisfies the so-called Kato cusp conditions at the coalescence point.\cite{kato57}
The family of methods which explicitly incorporate functions of the interelectron distance into the wave function ansatz to improve the description of the Coulomb hole 
are called explicitly correlated methods.\cite{hattig12,ten-no12,kong12}
With explicitly correlated methods, high accuracy can be reached using basis sets of moderate size (e.g. triple zeta). \newline
A prime example of explicitly correlated methods are the CC-F12 methods.\cite{klopper87,kutzelnigg91,klopper91,noga92,noga94,klopper02,valeev04,tenno04}
Without any approximations, already the explicitly correlated coupled cluster with singles and doubles (CCSD-F12) 
is too computationally expensive to be used for medium sized systems.\cite{tew10_rev,koehn08,shiozaki08a,shiozaki08b}
Many different approximations have been proposed that reduce the computational overhead of the F12 treatment.\cite{fliegl05,fliegl06,tew07,tew08,bokhan09,adler07,knizia09,valeev08a,valeev08b,torheyden09,haettig10}
The CC-F12x (x = a,b,*) methods are indeed only slightly more computationally demanding than the base CC method,
while retaining the faster convergence with respect to basis set size.\cite{adler07,knizia09,werner10_rev,haettig10}
But F12 methods reduce only the basis-set error and are difficult to extend to higher than double excitations.\cite{koehn09,koehn10} \newline
The transcorrelation approach \cite{boys69} to include explicit electron correlation 
offers a unique way to improve both the basis-set error and the method quality.
After its emergence in the sixties\cite{hirschfelder63,boys69,boys69_full,boys69_indeterminacy,boys69_bilinear,boys69_lih,handy69,handy71,handy72,handy73} 
the transcorrelation approach was dormant until the end of the century because of the difficult numerical calculation of the 
transcorrelated integrals, their daunting size and the non-Hermiticity of the transcorrelated two-electron integrals and the resulting loss of variationality.
Ideas by Nooijen and Bartlett in the late nineties\cite{nooijen98} and the frozen Gaussian geminal (FROGG) approach by Ten-no\cite{ten-no00,ten-no00_integrals} 
at the turn of the millennium have started a renewed interest in the transcorrelation approach and 
related similarity transformations of the Hamiltonian.\cite{hino02,imamura03,zweistra03,umezawa03,umezawa04_excited,umezawa04_ueg,umezawa05,yanai06,yanai07,neuscamman10,tsuneyuki08,
luo10,luo10_study,luo11,luo12,yanai12,sharma14,ochi12,ochi14,ochi15,yanai15,ochi16,wahlen-strothman15,kersten16,jeszenszki18,
luo18,dobrautz19,cohen19,guther21,liao21,schraivogel21_tc,luo22,motta20,baiardi20,baiardi22,khamoshi21,giner21,kumar22,dobrautz22,ammar22,liao22,ochiTC23,leeStudies2023,ammarBiorthonormal2023} 
A key recent finding has been that the non-Hermitian property of the TC method, long considered to be a nuisance, 
can actually be helpful in inducing compactness (and therefore ease of approximation) in the desired (right-eigenvector) solution, 
through the use of appropriate realistic Jastrow factors, 
substantially reducing the multi-configurational character of even strongly-correlated systems. 
This has been found for a variety of systems, including the 2D Hubbard model\cite{dobrautz19}, 3D uniform electron gas\cite{liao21}, 
and molecular systems\cite{schraivogel21_tc,liao22,haupt23}. This means that the TC method can improve the convergence of the {\em method}, and further evidence of this will be provided in this paper for a yet larger range of molecular systems. This effect goes substantially beyond the other popular form of quantum chemical explicit correlation, namely F12 methods, which improve basis-set convergence but do not accelerate convergence with respect to the method.
\chg{Although the F12 approach can be combined with multi-reference methods and thus is applicable to strongly multi-configurational problems,\cite{shiozaki_multireference_2013} it has not been found to reduce the strongly-correlated character of the problem.}  
\newline 
Another way to improve the CC results is to modify the amplitude 
equations\cite{paldus17,jankowski80,chiles81,paldus84,bartlett06,musial07,huntington10,huntington12,kats13,kats14,kats15,kats16,rishi17,kats18,kats19_dc,rishi19,schraivogel21_dc},
as has been done in the distinguishable cluster (DC) approximation\cite{kats13,kats14,kats15,kats16,rishi17,kats18,kats19_dc,rishi19,schraivogel21_dc}. 
The DC approximation has been developed to overcome the complete failure of coupled cluster with doubles (CCD) in systems with considerable amount of static electron correlation. 
But additionally, the DC approximation also improves absolute and relative energies of systems dominated by dynamical electron correlation. \newline
The transcorrelated Hamiltonian has already been combined with linearized CCSD by \mbox{Ten-no}, albeit using a considerably simpler Jastrow factor than in our work and 
including the three-electron integrals only approximately.\cite{hino02}
Some of the authors have investigated the uniform electron gas (UEG) over a wide range of densities with TC-CCD and transcorrelated distinguishable cluster with doubles (TC-DCD), 
while neglecting the effect of the explicit normal-ordered three-electron integrals.\cite{liao21} \newline
In a previous study on atoms, we used a sophisticated Jastrow factor and included all explicit three-electron integrals, 
resulting in full transcorrelated CCSD (TC-CCSD) and distinguishable cluster with singles and doubles (TC-DCSD) methods.\cite{schraivogel21_tc} 
Findings on the UEG have been replicated in the studied atoms as well, namely that the effect of the explicit three-electron integrals were negligible and 
two different approximations to the full TC treatment, called approximation~A and approximation~B have been
proposed and shown to capture the major effects of transcorrelation. \newline
We have improved our implementations of TC-CCSD and TC-DCSD with the help of automated implementation strategies, 
such that the methods can be applied to small molecules, and combined with the new Jastrow-factor optimization and 
efficient computation of transcorrelated integrals.\cite{haupt23} 
In this publication we report calculations of full TC-CCSD and TC-DCSD and their approximations on molecular systems.
\section{Theory}
\subsection{Transcorrelation}
Transcorrelated methods use the Jastrow factorization of the wave function $\Psi$
\begin{equation}
\Psi = e^{\tau} \Phi,
\label{eq:jastrow}
\end{equation}
with
\begin{equation}
\tau = \sum_{i \lt j}^{N} u(r_i,r_j),
\end{equation}
for a system with $N$ electrons and a real symmetric correlation factor $u(r_i,r_j)=u(r_j,r_i)$. \newline
Plugging equation~(\ref{eq:jastrow}) into the Schr\"odinger equation and multiplying with $e^{-\tau}$ from the left leads to a similarity transformed Hamiltonian $\bar H = e^{-\tau}\op H e^{\tau}$
with the corresponding Schr\"odinger equation
\begin{equation}
\bar H \Phi = E \Phi,
\end{equation}
with the original energy $E$ unaltered. \newline
Because the correlation function $\tau$ is not anti-Hermitian, the corresponding similarity transformation is not unitary. As a result the transcorrelated Hamiltonian is not Hermitian.
Furthermore, because a similarity transformation does not change the eigenvalues (in the hypothetical case of a complete basis),
there is in principle complete freedom in defining the functional form of $u(r_i,r_j)$.
But to actually improve the convergence with respect to basis set size and method level, it should be chosen such that $\Phi$ is a smoother and less complicated wave function than $\Psi$.
Often it is chosen to fulfill Kato's cusp condition\cite{kato57} itself\cite{schmidt90,giner21}
or to ensure that the resulting transcorrelated wave function (approximately) fulfills the condition\cite{ten-no00}. 
The Baker-Campbell-Hausdorff expansion of the similarity transformed Hamiltonian truncates after the second commutator without any approximation, because
$u$ commutes with every operator in the Hamiltonian except the kinetic energy operator and every commutator reduces the differentiation order by one.
From the second commutator a three-electron integral arises,
\begin{equation}
\sum_n (\nabla_{n} \tau)^{2} = \sum_{nlq} (\nabla_{n} u_{nl}) \cdot (\nabla_{n} u_{nq}),
\end{equation}
which can considerably increase the computational cost of transcorrelated methods.
However, as demonstrated in our previous publications\cite{liao21,schraivogel21_tc} and this contribution, taking into account only the mean-field effect 
of the three-electron integrals results in negligible errors.
The Jastrow factor is conveniently optimized in first quantization and can be system-specific or universal\cite{ten-no00}.
The Jastrow ansatz leads to a similarity transformed Hamiltonian, which can be written in second quantization as,
\begin{align}
&\bar H = \sum_{pq}h_p^q \sum_{\sigma} a^\dg_{p\sigma}a_{q \sigma} +
\half\sum_{pqrs}\lk(V_{pr}^{qs} - K_{pr}^{qs}\rk)\sum_{\sigma\rho}a^\dg_{p\sigma} a^\dg_{r\rho} a_{s \rho} a_{q \sigma}\nl
-\frac{1}{6}\sum_{pqrstu}L_{prt}^{qsu}\sum_{\sigma\rho\tau}a^\dg_{p\sigma} a^\dg_{r\rho}a^\dg_{t\tau} a_{u \tau} a_{s \rho} a_{q \sigma},
\label{eq:tch}
\end{align}
with $h_p^q = < p | \op h | q >$ and $V_{pr}^{qs} = <pr|qs>$ being the one and two electron integrals of the original electronic Hamiltonian and the non-Hermitian
two-electron integral $K$ and the three-electron integral $L$ with full 48-fold symmetry for real orbitals coming from the similarity transformation of the Hamiltonian with $e^{\tau}$.
\subsection{Transcorrelated Coupled Cluster}
The lowest eigenvalue of $\bar H$, which corresponds to the ground-state energy, can be calculated with any method that can be applied to non-Hermitian Hamiltonians.
Here we are employing coupled cluster methods as a convenient alternative to FCI.
The additional contributions to the amplitude and energy equation of CC can be derived using standard second quantization techniques.
The contribution to the energy $E_{\mathrm{TC}}$ is given as,
\begin{equation}
E_{\mathrm{TC}} = L_{ikm}^{jln}\left( -\frac{1}{6} \delta_{ij}\delta_{kl}\delta_{mn} + \frac{1}{2} \delta_{ij}\delta_{kn}\delta_{ml} 
                   - \frac{1}{3} \delta_{in}\delta_{kj}\delta_{ml} \right),
\end{equation}
with occupied spin-orbital indices $(i,j,..)$. \newline
Contributions of the three-body integrals without self-contractions have to be added explicitly to the singles and doubles amplitude equations.
For both proposed approximations (see below) those contributions are neglected.
The remaining (and more important) contributions to the energy and amplitude equations are captured by
contracting the three-body integrals at the beginning of every CC iteration on the two-body integrals as, 
\begin{equation}\label{eq:2bodytc}
(pq|rs) \mathrel{-}\mathbin{=} \left( L_{prk}^{qsl} - L_{prk}^{qls} - L_{prk}^{lsq} \right)\delta_{kl},
\end{equation}
with the indices $p,q,r,s$ going over the complete spin-orbital space. Note that the one body contribution from the modified two-body integrals requires corrections in order to remove the double counting of TC integrals. \newline
Using integrals dressed with singles-amplitudes significantly simplifies the TC-CC amplitude equations.\cite{koch94}
The \mbox{TC-CCSD} method scales as $\mathcal{O}(N^7)$, with $N$ being a measure of the system size, and dealing with the full $L$ tensor is memory and I/O intensive.
In our previous publication we proposed approximations to the full TC-CCSD method based on normal-order theory, 
called approximation~A (TC-CCSD-A) and approximation~B (TC-CCSD-B).\cite{schraivogel21_tc}
Approximation~A neglects three-body integrals normal-ordered with respect to a "Br\"uckner optimized determinant", i.e. the Hartree-Fock determinant that has been rotated with the CC singles amplitudes.
The more economical approximation~B avoids the expensive dressing of the three-electron integrals with the singles amplitudes and 
neglects three-body integrals normal-ordered with respect to the Hartree-Fock determinant and 
as the result has the same nominal scaling of $\mathcal{O}(N^6)$ as canonical CCSD.
For approximation~B the contraction in equation~\ref{eq:2bodytc} is only done once before the iterative solution of the CC equations.

The distinguishable cluster approximation can be applied to the TC-CCSD amplitude equations in the usual way, resulting in the TC-DCSD method.
We would like to stress that approximation~B leads in practice to an effective Hamiltonian with no more than two-body interactions.
Standard implementations of quantum chemistry methods, including coupled cluster methods with arbitrarily high excitations, can be used to obtain the ground-state of the effective Hamiltonian, 
if the non-Hermiticity of the two-body integrals is taken into account.
\section{Results}
The TC-CCSD and TC-DCSD spin-factored equations were derived, optimized and implemented using our second-quantization program Quantwo\cite{quantwo} and the
Integrated Tensor Framework (ITF)\cite{shamasundar2011,MOLPRO-WIREs} in \molpro\cite{MOLPRO-WIREs, MOLPRO-JCP, MOLPRO}.
The calculations have been done using the cc-pVTZ basis set\cite{dunning89} and all electrons have been correlated.
We used the following form of the correlation factor,\cite{foulkes01,drummond04}
\begin{equation}
    \hat{\tau} = \sum_{i \gt j}u(r_{ij})+\sum_{I=1}^{N_{\rm atoms}}\sum_{i=1}^{N}\chi_I(r_{iI})+\sum_{I=1}^{N_{\rm atoms}}\sum^N_{i \gt j}f_I(r_{iI}, r_{jI}, r_{ij}),
\end{equation}
with electron-electron terms~$u$, electron-nucleus terms~$\chi$ and electron-electron-nucleus terms~$f$ for a system with 
$N$~electrons with electron indices~$i$~and~$j$ and $N_{\rm atoms}$ with index~$I$.
Spin independent Jastrow factors have been used throughout this work, and the cutoff lengths have been fixed to 4 and 4.5 bohr for the electron-electron and atom-centered terms, respectively. The correlation factor has been optimized by minimizing the variance\cite{umrigar88,kent99} of the reference energy
within variational Monte Carlo (VMC) using the \textsc{casino} program\cite{needs20}; the nominal scaling of VMC is $\mathcal{O}(N^3)$.
All results presented in this publication have been obtained using Jastrow 
factors optimized with a single determinant (Hartree-Fock) reference function.
The transcorrelated integrals have been calculated using the TCHInt library as outlined in Ref.~\onlinecite{cohen19}, 
which uses numerical quadrature with grids provided by PySCF\cite{sun15,sun18,sun20} to calculate the transcorrelated integrals.
More details about the Jastrow optimization can be found in Ref.~\onlinecite{haupt23}.
The F12 results have been obtained using unrestricted CCSD-F12a, DCSD-F12a,
and CCSD(T)-F12a methods\cite{knizia09,kats15} \chg{with the diagonal ansatz 3C(D)\cite{kedzuch05, werner07}}, as implemented in \molpro.
Orbitals in all our calculations have been obtained using R(O)HF.
\subsection{Absolute and atomization energies}
\begin{figure}
  \centering
  \includegraphics[scale=1.0]{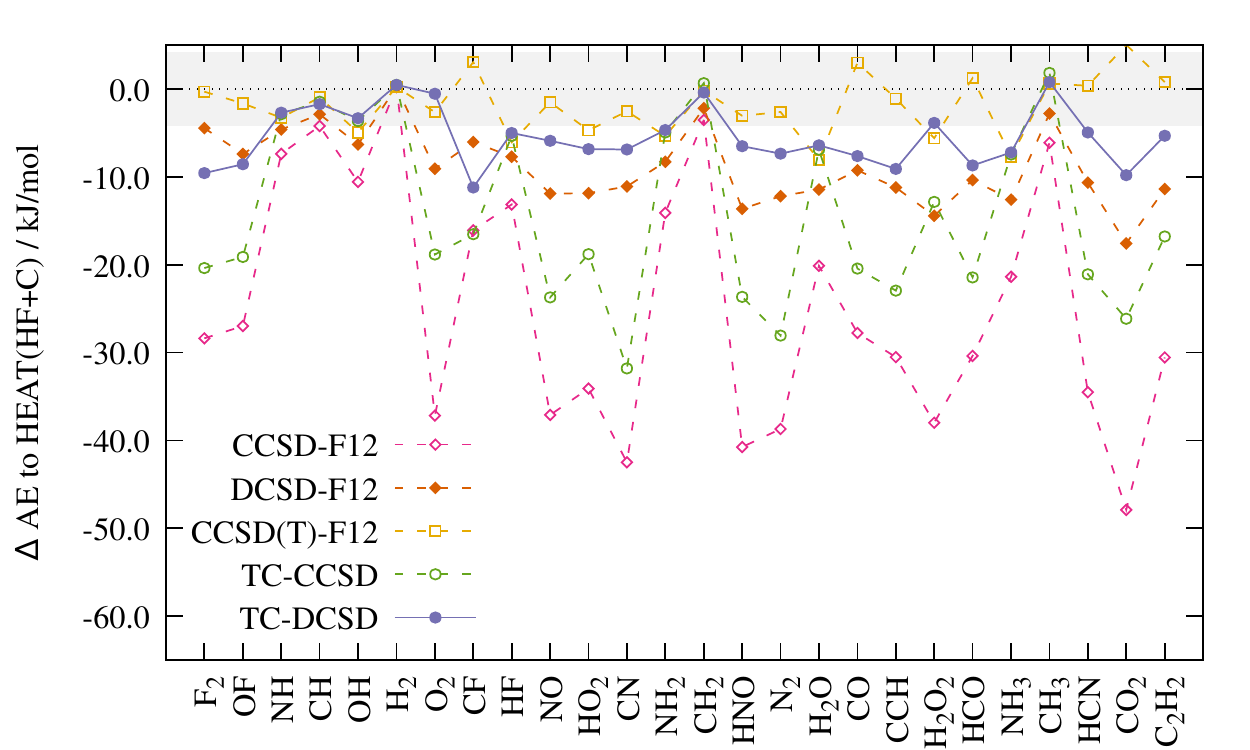}
  \caption{The difference between calculated cc-pVTZ atomization energies with TC-CCSD, TC-DCSD and various CC-F12 methods compared to HEAT(HF+C) is shown.
  The region of chemical accuracy is shaded.}
  \label{fig:fulltc}
\end{figure}

\begin{figure}
  \centering
  \includegraphics[scale=1.0]{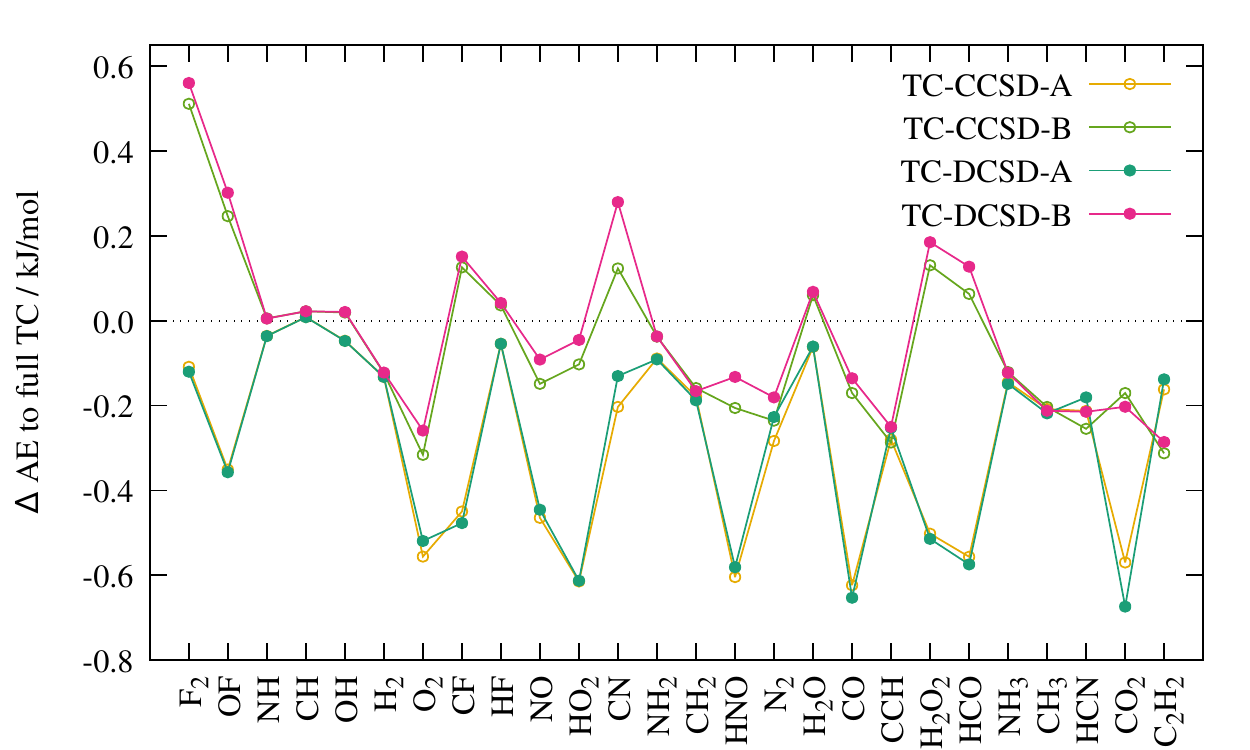}
  \caption{The difference between calculated cc-pVTZ atomization energies with the full transcorrelated methods and the two tested approximations is shown.}
  \label{fig:approxtc}
\end{figure}

\begin{figure}
  \centering
  \includegraphics[scale=1.0]{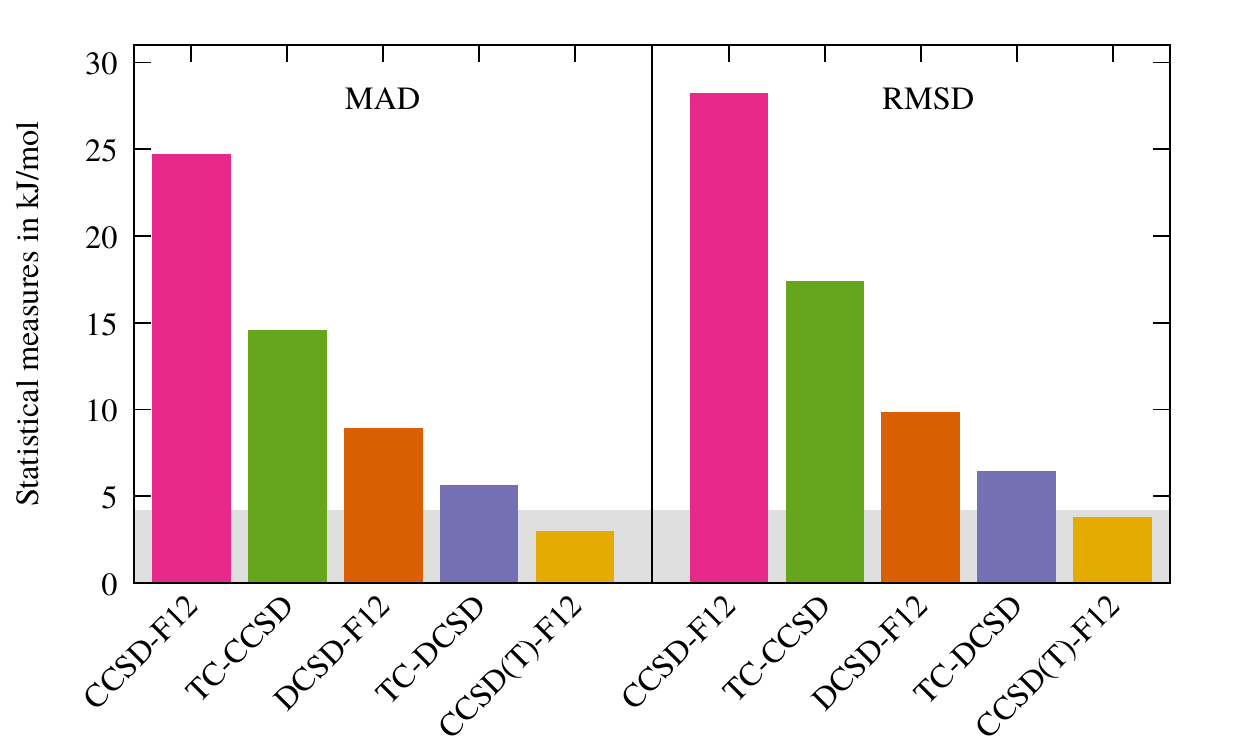}
  \caption{The mean absolute deviation (MAD) and the root-mean-squared deviation (RMSD) of the calculated cc-pVTZ atomization energies 
           relative to the HEAT(HF+C) values are shown. The region of chemical accuracy is shaded.}
  \label{fig:heat_stats}
\end{figure}

We benchmarked the TC-CC methods against highly accurate reference values from the literature taken from 
the ``high-accuracy extrapolated ab-initio thermochemistry'' (HEAT) project.\cite{tajti04,bomble06,harding08,thorpe19}
The reference energies were taken from Ref.~\onlinecite{tajti04}, where they have been calculated as,
\begin{equation}
E_{\mathrm{HEAT}}^{\mathrm{HF+C}} = E^{\infty}_{\mathrm{HF}} + \Delta E^{\infty}_{\mathrm{CCSD(T)}} + \Delta E^{\mathrm{TQ}}_{\mathrm{CCSDT}}(\mathrm{fc}) + E^{\mathrm{cc-pVDZ}}_{\mathrm{CCSDTQ}}(\mathrm{fc}),
\end{equation}
with $E^{\infty}_{\mathrm{HF}}$ an aug-cc-pCVXZ(X=T,Q,5) extrapolation\cite{feller92,feller93}, $\Delta E^{\infty}_{\mathrm{CCSD(T)}}$ an aug-cc-pCVXZ(X=Q,5) extrapolation\cite{helgaker97}, 
$\Delta E^{\mathrm{TQ}}_{\mathrm{CCSDT}}(\mathrm{fc})$ an cc-pVXZ(X=T,Q) extrapolation using the frozen-core (fc) approximation and no extrapolation for the $E^{\mathrm{cc-pVDZ}}_{\mathrm{CCSDTQ}}(\mathrm{fc})$
contribution has been used.
The overall performance of the transcorrelated CC methods for the atomization energies on the HEAT set is presented in Fig.~\ref{fig:fulltc}.
The quality of the mean-field approximations is exemplified with the atomization energies in Fig.~\ref{fig:approxtc}.
The accompanying statistics including the absolute energies of
Fig.~\ref{fig:fulltc} and  Fig.~\ref{fig:approxtc} are shown in Table~\ref{tab:heat_stats} 
and the statistics of Table~\ref{tab:heat_stats} are visualized for the atomization energies in Fig.~\ref{fig:heat_stats}.
\begin{table*}[htbp]
\caption{Statistics for cc-pVTZ absolute and relative energies compared to HEAT(HF+C).}
\label{tab:heat_stats}
\begin{ruledtabular}
\begin{tabular}{lllllll}
             & \multicolumn{3}{l}{absolute energies / millihartree} & \multicolumn{3}{l}{atomization energies / kJ/mol}      \\
            & MAD   & SD                               & MaxD   & MAD   & RMSD                          & MaxD   \\
CCSD-F12    &  \chg{32.60} & \chg{18.01} & \chg{68.43} & \chg{24.69} & \chg{28.17} & \chg{-47.91} \\
DCSD-F12    &  \chg{24.35} & \chg{12.86} & \chg{50.34} & \chg{8.90}  & \chg{9.83}  & \chg{-17.57} \\
CCSD(T)-F12 &  \chg{21.35} & \chg{11.12} & \chg{43.09} & \chg{2.94}  & \chg{3.73}  & \chg{-8.07}  \\
TC-CCSD      & 17.53 & 11.19 & 39.97 & 14.54 & 17.37 & -31.80 \\
TC-CCSD-A    & 17.48 & 11.17 & 40.03 & 14.29 & 17.06 & -31.60 \\
TC-CCSD-B    & 17.56 & 11.24 & 40.25 & 14.52 & 17.32 & -31.93 \\
TC-DCSD      & 13.68 & 9.09  & 35.13 & 5.59  & 6.38  & -11.21 \\
TC-DCSD-A    & 13.64 & 9.07  & 35.20 & 5.33  & 6.10  & -10.74 \\
TC-DCSD-B    & 13.72 & 9.15  & 35.43 & 5.59  & 6.40  & -11.37
\end{tabular}
\end{ruledtabular}
\end{table*}
The transcorrelated CC calculations are more accurate than the F12 values for the absolute energies. 
Furthermore, the difference between the full transcorrelated CC and their approximations are very small for the absolute and atomization
energies, with maximal deviations of around 0.6~kJ/mol for the atomization energies.
For the absolute and atomization energies, the TC-CC results are noticeably better than the corresponding F12 values of the same CC method.
The root-mean-squared deviation (RMSD) of the atomization energies calculated with CCSD-F12 of \chg{28.17} kJ/mol is almost halved by TC-CCSD to 17.37 kJ/mol.
Moreover, the DC approximation applied to the TC-CCSD method significantly improves the results, especially the atomization energies, 
bringing the RMSD close to chemical accuracy from 17.37 to 6.38 kJ/mol.
However, the atomization energies from CCSD(T)-F12 are still closer to the HEAT(HF+C) values, for the price of a nominal $\mathcal{O}(N^7)$ scaling, 
compared to the $\mathcal{O}(N^6)$ scaling of TC-DCSD-B. \newline
\subsection{Dissociation of the nitrogen molecule}
\begin{figure}
  \centering
  \includegraphics[scale=1.0]{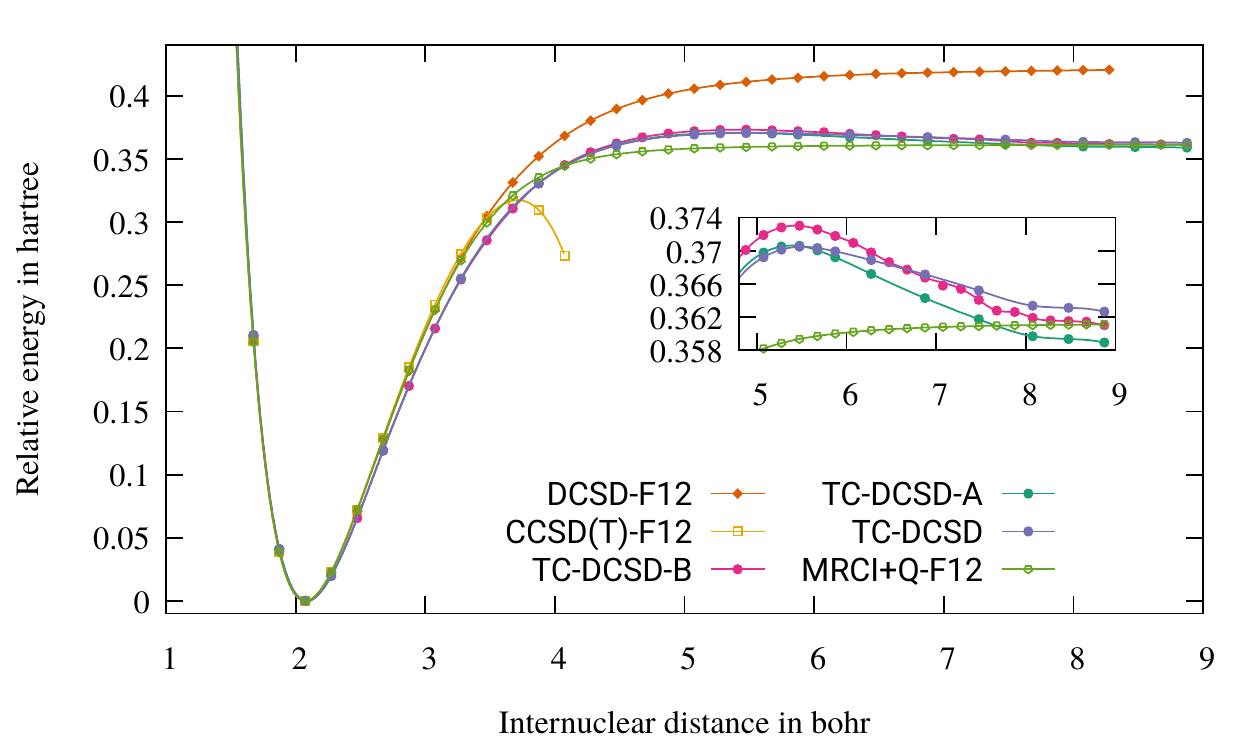}
  \caption{The difference between the full transcorrelated TC-DCSD and the two tested approximations is monitored during the dissociation of the nitrogen molecule. The cc-pVTZ basis has been used.}
  \label{fig:n2}
\end{figure}

In this section we investigate, as a prototypical test case spanning several different correlation regimes, 
the dissociation of the nitrogen molecule with the TC-CC methods.
At equilibrium the wave function of the nitrogen molecule is dominated by a single Slater determinant, 
whose importance in the wave function quickly diminishes and many determinants with equal weight become
important along the dissociation path.
That makes the breaking of the nitrogen molecule a challenge for quantum chemistry methods.
At stretched geometries, CC calculations are complicated by the existence of several close-lying solutions.
To remedy this problem the amplitudes of the previous geometry have been used as a starting guess as the bond is stretched in successive calculations.
The MRCI+Q-F12 reference values have been calculated using \molpro\cite{WK88,KW88,Shiozaki:2011,Shiozaki:2011a,shiozaki_multireference_2013}.
The calculated energies along the dissociation of the nitrogen molecule are shown in Figure~\ref{fig:n2}.
The potential energy curve is described overall extremely well by TC-DCSD and the two approximations.
Almost perfect agreement is observed for every method around the equilibrium and at compressed geometries.
When the bond is stretched the TC-DCSD method and the two approximations are up to ten millihartree below the reference curve.
But even at stretched geometries the mean-field approximations work extremely well.
At the beginning of the spin-recoupling region around 3~bohr the F12 methods are closer to the reference than the TC methods.
However, both F12 methods show the well known deficiencies at dissociation.
CCSD(T)-F12 breaks down completely, while DCSD-F12 breaks the bond qualitatively correctly, but overbinds the molecule severely.
After a crossover region with the reference curve at 4~bohr a small bump is observed for the TC methods, which is not there in the original DCSD method.\cite{kats13,kats14} 
Thus, we relate this deficiency to the Jastrow factor optimization, and 
expect that it will disappear if multi-determinant reference functions are employed 
in the VMC Jastrow optimisation procedure, for such highly multi-configurational systems.
Nevertheless, in contrast to DCSD, TC-DCSD does not overbind the molecule and converges to the correct dissociation limit.
Albeit small, the differences between TC-DCSD and the two approximations in the spin-recoupling region are bigger than everything we have seen in our benchmarks
on the HEAT set.
The predicted dissociation energies of the nitrogen molecule by the TC methods are astonishingly good 
for single-reference CC methods and are listed in Table~\ref{tab:de} together with the theoretical reference value 
and the DCSD-F12 result. 
For perspective, the aug-cc-pV5Z dissociation energy calculated with MRCI+Q-F12 is 9.90~eV.
\begin{table*}[htpb]
\caption{Calculated dissociation energies (cc-pVTZ) of the nitrogen molecule in eV.}
\begin{ruledtabular}
\begin{tabular}{lllll}
TC-DCSD & TC-DCSD-A & TC-DCSD-B & DCSD-F12  &MRCI+Q-F12  \\
 9.87    & 9.77      & 9.82    & 11.44     &9.83  
\end{tabular}
\end{ruledtabular}
\label{tab:de}
\end{table*}
\section{Conclusions}
The new Jastrow optimization technique\cite{haupt23} combined with the transcorrelated coupled cluster and distinguishable cluster methods yield
very accurate absolute and relative energies for molecules.
The approximation of neglecting the normal-ordered explicit three-body integrals has been shown to introduce only very small errors for absolute and relative energies
for a variety of different atoms and molecules from the HEAT set and different correlation regimes during the dissociation of the nitrogen molecule.
It was demonstrated that the TC-DCSD methods yield highly accurate dissociation energies for the nitrogen molecule, 
which is remarkable considering that we still operate conceptually in a single-reference CC framework.
\chg{However, more work is required to improve the TC-DCSD accuracy in the intermediate stretched regime.}
When the HF determinant is not a good approximation to the total wave function, larger deviations between the full TC and the approximated versions are to be expected.
For example, during the dissociation of the nitrogen molecule deviations up to 2.5 millihartree at the spin-recoupling region have been observed 
for our most practical approximation~B from the full TC-DCSD.
We have shown that absolute and relative energies calculated with TC-CCSD and TC-DCSD are much closer to the reference values than the corresponding F12 methods 
and the error from the mean-field approximation is negligible.
At the cc-pVTZ level CCSD(T)-F12 calculated more accurate atomization energies with a RMSD of \chg{3.73} kJ/mol, 
compared to our most economical approximation~B of TC-DCSD (RMSD of 6.10 kJ/mol) for the price of a nominal higher cost of $\mathcal{O}(N^7)$, 
in contrast to the $\mathcal{O}(N^6)$ cost of TC-DCSD-B.
The TC-CCSD method outperformed CCSD-F12 for absolute and relative energies on the HEAT set.
However, as known for F12 methods, diffuse basis sets are necessary for high accuracy.\cite{tew05}
The effect of basis sets for TC methods will be investigated in a forthcoming publication, 
together with a very efficient implementation of approximation~B.
We would like to emphasize that approximation~B is quite general and can be combined with various many-body methods  
that can handle non-Hermitian Hamiltonians, including coupled cluster with arbitrarily high excitations.
Furthermore, it can also be used to calculate excited states.\cite{liao22}
We are currently investigating methods to improve the Jastrow factors, such as incorporation of spin-dependency, as well as their 
optimization using multi-determinantal reference functions.
\section*{Supplementary material}
See the supplementary material for cc-pVTZ absolute and atomization energies of systems in the HEAT set
and the absolute energies of the dissociation curve of the nitrogen molecule calculated with the presented methods.
\chg{We also list F12 calculations with the often used fixed amplitude ansatz there and F12 calculations using the cc-pwCVTZ basis set.}

\begin{acknowledgements}
Funded by the Deutsche Forschungsgemeinschaft (DFG, German Research Foundation) -- 455145945. Financial support from the Max-Planck Society is gratefully acknowledged.
P.L.R. and A.A. acknowledge support from the European Centre of
Excellence in Exascale Computing TREX, funded by the Horizon 2020
program of the European Union under grant no.\ 952165.
Any views and opinions expressed are those of the authors only and do
not necessarily reflect those of the European Union or the European
Research Executive Agency.
Neither the European Union nor the granting authority can be held
responsible for them.
\end{acknowledgements}

\section*{Author Declarations}
The authors have no conflicts to disclose.

\section*{Author contributions}
\textbf{Thomas Schraivogel:} Data curation (lead); Investigation (equal); Software (equal); Writing/Original Draft Preparation (lead); Writing/Review and Editing (equal). \textbf{Evelin Martine Christlmaier:} Software (equal); Writing/Review and Editing (equal); Investigation (equal). \textbf{Pablo L\'opez R\'ios:} Software (equal); Writing/Review and Editing (equal). \textbf{Ali Alavi:} Funding acquisition (equal); Resources (lead); Writing/Review and Editing (equal); Conceptualization (equal); Supervision (supporting). \textbf{Daniel Kats:} Conceptualization (equal); Funding acquisition (equal); Writing/Review and Editing (equal); Software (equal); Supervision (lead).

\section*{Data availability}
The data that support the findings of this study are available within the article and its supplementary material.

\bibliography{books,molpro,dc,f12,qmc,tc,software,thermochem,cc,qc,cc_internal,automation}

\begin{thebibliography}{141}%
\makeatletter
\providecommand \@ifxundefined [1]{%
 \@ifx{#1\undefined}
}%
\providecommand \@ifnum [1]{%
 \ifnum #1\expandafter \@firstoftwo
 \else \expandafter \@secondoftwo
 \fi
}%
\providecommand \@ifx [1]{%
 \ifx #1\expandafter \@firstoftwo
 \else \expandafter \@secondoftwo
 \fi
}%
\providecommand \natexlab [1]{#1}%
\providecommand \enquote  [1]{``#1''}%
\providecommand \bibnamefont  [1]{#1}%
\providecommand \bibfnamefont [1]{#1}%
\providecommand \citenamefont [1]{#1}%
\providecommand \href@noop [0]{\@secondoftwo}%
\providecommand \href [0]{\begingroup \@sanitize@url \@href}%
\providecommand \@href[1]{\@@startlink{#1}\@@href}%
\providecommand \@@href[1]{\endgroup#1\@@endlink}%
\providecommand \@sanitize@url [0]{\catcode `\\12\catcode `\$12\catcode
  `\&12\catcode `\#12\catcode `\^12\catcode `\_12\catcode `\%12\relax}%
\providecommand \@@startlink[1]{}%
\providecommand \@@endlink[0]{}%
\providecommand \url  [0]{\begingroup\@sanitize@url \@url }%
\providecommand \@url [1]{\endgroup\@href {#1}{\urlprefix }}%
\providecommand \urlprefix  [0]{URL }%
\providecommand \Eprint [0]{\href }%
\providecommand \doibase [0]{http://dx.doi.org/}%
\providecommand \selectlanguage [0]{\@gobble}%
\providecommand \bibinfo  [0]{\@secondoftwo}%
\providecommand \bibfield  [0]{\@secondoftwo}%
\providecommand \translation [1]{[#1]}%
\providecommand \BibitemOpen [0]{}%
\providecommand \bibitemStop [0]{}%
\providecommand \bibitemNoStop [0]{.\EOS\space}%
\providecommand \EOS [0]{\spacefactor3000\relax}%
\providecommand \BibitemShut  [1]{\csname bibitem#1\endcsname}%
\let\auto@bib@innerbib\@empty
\bibitem [{\citenamefont {\v{C}{\'i}\v{z}ek}(1966)}]{cizek66}%
  \BibitemOpen
  \bibfield  {author} {\bibinfo {author} {\bibfnamefont {J.}~\bibnamefont
  {\v{C}{\'i}\v{z}ek}},\ }\href {\doibase 10.1063/1.1727484} {\bibfield
  {journal} {\bibinfo  {journal} {J. Chem. Phys.}\ }\textbf {\bibinfo {volume}
  {45}},\ \bibinfo {pages} {4256} (\bibinfo {year} {1966})}\BibitemShut
  {NoStop}%
\bibitem [{\citenamefont {Purvis}\ and\ \citenamefont
  {Bartlett}(1982)}]{purvis82}%
  \BibitemOpen
  \bibfield  {author} {\bibinfo {author} {\bibfnamefont {G.~D.}\ \bibnamefont
  {Purvis}}\ and\ \bibinfo {author} {\bibfnamefont {R.~J.}\ \bibnamefont
  {Bartlett}},\ }\href {\doibase 10.1063/1.443164} {\bibfield  {journal}
  {\bibinfo  {journal} {J. Chem. Phys.}\ }\textbf {\bibinfo {volume} {76}},\
  \bibinfo {pages} {1910} (\bibinfo {year} {1982})}\BibitemShut {NoStop}%
\bibitem [{\citenamefont {Bartlett}\ and\ \citenamefont
  {Musia\l{}}(2007)}]{bartlett07}%
  \BibitemOpen
  \bibfield  {author} {\bibinfo {author} {\bibfnamefont {R.~J.}\ \bibnamefont
  {Bartlett}}\ and\ \bibinfo {author} {\bibfnamefont {M.}~\bibnamefont
  {Musia\l{}}},\ }\href {\doibase 10.1103/RevModPhys.79.291} {\bibfield
  {journal} {\bibinfo  {journal} {Rev. Mod. Phys.}\ }\textbf {\bibinfo {volume}
  {79}},\ \bibinfo {pages} {291} (\bibinfo {year} {2007})}\BibitemShut
  {NoStop}%
\bibitem [{\citenamefont {Kucharski}\ and\ \citenamefont
  {Bartlett}(1992)}]{kucharski92}%
  \BibitemOpen
  \bibfield  {author} {\bibinfo {author} {\bibfnamefont {S.~A.}\ \bibnamefont
  {Kucharski}}\ and\ \bibinfo {author} {\bibfnamefont {R.~J.}\ \bibnamefont
  {Bartlett}},\ }\href {\doibase 10.1063/1.463930} {\bibfield  {journal}
  {\bibinfo  {journal} {J. Chem. Phys.}\ }\textbf {\bibinfo {volume} {97}},\
  \bibinfo {pages} {4282} (\bibinfo {year} {1992})}\BibitemShut {NoStop}%
\bibitem [{\citenamefont {Bomble}\ \emph {et~al.}(2005)\citenamefont {Bomble},
  \citenamefont {Stanton}, \citenamefont {Kállay},\ and\ \citenamefont
  {Gauss}}]{bomble05}%
  \BibitemOpen
  \bibfield  {author} {\bibinfo {author} {\bibfnamefont {Y.~J.}\ \bibnamefont
  {Bomble}}, \bibinfo {author} {\bibfnamefont {J.~F.}\ \bibnamefont {Stanton}},
  \bibinfo {author} {\bibfnamefont {M.}~\bibnamefont {Kállay}}, \ and\
  \bibinfo {author} {\bibfnamefont {J.}~\bibnamefont {Gauss}},\ }\href
  {\doibase 10.1063/1.1950567} {\bibfield  {journal} {\bibinfo  {journal} {J.
  Chem. Phys.}\ }\textbf {\bibinfo {volume} {123}},\ \bibinfo {pages} {054101}
  (\bibinfo {year} {2005})}\BibitemShut {NoStop}%
\bibitem [{\citenamefont {Tajti}\ \emph {et~al.}(2004)\citenamefont {Tajti},
  \citenamefont {Szalay}, \citenamefont {Cs\'asz\'ar}, \citenamefont
  {K\'allay}, \citenamefont {Gauss}, \citenamefont {Valeev}, \citenamefont
  {Flowers}, \citenamefont {V\'azquez},\ and\ \citenamefont
  {Stanton}}]{tajti04}%
  \BibitemOpen
  \bibfield  {author} {\bibinfo {author} {\bibfnamefont {A.}~\bibnamefont
  {Tajti}}, \bibinfo {author} {\bibfnamefont {P.~G.}\ \bibnamefont {Szalay}},
  \bibinfo {author} {\bibfnamefont {A.~G.}\ \bibnamefont {Cs\'asz\'ar}},
  \bibinfo {author} {\bibfnamefont {M.}~\bibnamefont {K\'allay}}, \bibinfo
  {author} {\bibfnamefont {J.}~\bibnamefont {Gauss}}, \bibinfo {author}
  {\bibfnamefont {E.~F.}\ \bibnamefont {Valeev}}, \bibinfo {author}
  {\bibfnamefont {B.~A.}\ \bibnamefont {Flowers}}, \bibinfo {author}
  {\bibfnamefont {J.}~\bibnamefont {V\'azquez}}, \ and\ \bibinfo {author}
  {\bibfnamefont {J.~F.}\ \bibnamefont {Stanton}},\ }\href {\doibase
  10.1063/1.1811608} {\bibfield  {journal} {\bibinfo  {journal} {J. Chem.
  Phys.}\ }\textbf {\bibinfo {volume} {121}},\ \bibinfo {pages} {11599}
  (\bibinfo {year} {2004})}\BibitemShut {NoStop}%
\bibitem [{\citenamefont {Kato}(1957)}]{kato57}%
  \BibitemOpen
  \bibfield  {author} {\bibinfo {author} {\bibfnamefont {T.}~\bibnamefont
  {Kato}},\ }\href@noop {} {\bibfield  {journal} {\bibinfo  {journal} {Comm.
  Pure Appl. Math.}\ }\textbf {\bibinfo {volume} {10}},\ \bibinfo {pages} {151}
  (\bibinfo {year} {1957})}\BibitemShut {NoStop}%
\bibitem [{\citenamefont {H\"attig}\ \emph {et~al.}(2012)\citenamefont
  {H\"attig}, \citenamefont {Klopper}, \citenamefont {K\"ohn},\ and\
  \citenamefont {Tew}}]{hattig12}%
  \BibitemOpen
  \bibfield  {author} {\bibinfo {author} {\bibfnamefont {C.}~\bibnamefont
  {H\"attig}}, \bibinfo {author} {\bibfnamefont {W.}~\bibnamefont {Klopper}},
  \bibinfo {author} {\bibfnamefont {A.}~\bibnamefont {K\"ohn}}, \ and\ \bibinfo
  {author} {\bibfnamefont {D.~P.}\ \bibnamefont {Tew}},\ }\href {\doibase
  10.1021/cr200168z} {\bibfield  {journal} {\bibinfo  {journal} {Chem. Rev.}\
  }\textbf {\bibinfo {volume} {112}},\ \bibinfo {pages} {4} (\bibinfo {year}
  {2012})}\BibitemShut {NoStop}%
\bibitem [{\citenamefont {Ten-no}(2012)}]{ten-no12}%
  \BibitemOpen
  \bibfield  {author} {\bibinfo {author} {\bibfnamefont {S.}~\bibnamefont
  {Ten-no}},\ }\href {\doibase 10.1007/s00214-011-1070-1} {\bibfield  {journal}
  {\bibinfo  {journal} {Theor. Chim. Acta}\ }\textbf {\bibinfo {volume}
  {131}},\ \bibinfo {pages} {1070} (\bibinfo {year} {2012})}\BibitemShut
  {NoStop}%
\bibitem [{\citenamefont {Kong}, \citenamefont {Bischoff},\ and\ \citenamefont
  {Valeev}(2012)}]{kong12}%
  \BibitemOpen
  \bibfield  {author} {\bibinfo {author} {\bibfnamefont {L.}~\bibnamefont
  {Kong}}, \bibinfo {author} {\bibfnamefont {F.~A.}\ \bibnamefont {Bischoff}},
  \ and\ \bibinfo {author} {\bibfnamefont {E.~F.}\ \bibnamefont {Valeev}},\
  }\href {\doibase 10.1021/cr200204r} {\bibfield  {journal} {\bibinfo
  {journal} {Chem. Rev.}\ }\textbf {\bibinfo {volume} {112}},\ \bibinfo {pages}
  {75} (\bibinfo {year} {2012})}\BibitemShut {NoStop}%
\bibitem [{\citenamefont {Klopper}\ and\ \citenamefont
  {Kutzelnigg}(1987)}]{klopper87}%
  \BibitemOpen
  \bibfield  {author} {\bibinfo {author} {\bibfnamefont {W.}~\bibnamefont
  {Klopper}}\ and\ \bibinfo {author} {\bibfnamefont {W.}~\bibnamefont
  {Kutzelnigg}},\ }\href {\doibase 10.1016/0009-2614(87)80005-2} {\bibfield
  {journal} {\bibinfo  {journal} {Chem. Phys. Lett.}\ }\textbf {\bibinfo
  {volume} {134}},\ \bibinfo {pages} {17} (\bibinfo {year} {1987})}\BibitemShut
  {NoStop}%
\bibitem [{\citenamefont {Kutzelnigg}\ and\ \citenamefont
  {Klopper}(1991)}]{kutzelnigg91}%
  \BibitemOpen
  \bibfield  {author} {\bibinfo {author} {\bibfnamefont {W.}~\bibnamefont
  {Kutzelnigg}}\ and\ \bibinfo {author} {\bibfnamefont {W.}~\bibnamefont
  {Klopper}},\ }\href {\doibase 10.1063/1.459921} {\bibfield  {journal}
  {\bibinfo  {journal} {J. Chem. Phys.}\ }\textbf {\bibinfo {volume} {94}},\
  \bibinfo {pages} {1985} (\bibinfo {year} {1991})}\BibitemShut {NoStop}%
\bibitem [{\citenamefont {Klopper}(1991)}]{klopper91}%
  \BibitemOpen
  \bibfield  {author} {\bibinfo {author} {\bibfnamefont {W.}~\bibnamefont
  {Klopper}},\ }\href@noop {} {\bibfield  {journal} {\bibinfo  {journal} {Chem.
  Phys. Lett.}\ }\textbf {\bibinfo {volume} {186}},\ \bibinfo {pages} {583}
  (\bibinfo {year} {1991})}\BibitemShut {NoStop}%
\bibitem [{\citenamefont {Noga}, \citenamefont {Kutzelnigg},\ and\
  \citenamefont {Klopper}(1992)}]{noga92}%
  \BibitemOpen
  \bibfield  {author} {\bibinfo {author} {\bibfnamefont {J.}~\bibnamefont
  {Noga}}, \bibinfo {author} {\bibfnamefont {W.}~\bibnamefont {Kutzelnigg}}, \
  and\ \bibinfo {author} {\bibfnamefont {W.}~\bibnamefont {Klopper}},\
  }\href@noop {} {\bibfield  {journal} {\bibinfo  {journal} {Chem. Phys.
  Lett.}\ }\textbf {\bibinfo {volume} {199}},\ \bibinfo {pages} {497} (\bibinfo
  {year} {1992})}\BibitemShut {NoStop}%
\bibitem [{\citenamefont {Noga}\ and\ \citenamefont
  {Kutzelnigg}(1994)}]{noga94}%
  \BibitemOpen
  \bibfield  {author} {\bibinfo {author} {\bibfnamefont {J.}~\bibnamefont
  {Noga}}\ and\ \bibinfo {author} {\bibfnamefont {W.}~\bibnamefont
  {Kutzelnigg}},\ }\href@noop {} {\bibfield  {journal} {\bibinfo  {journal} {J.
  Chem. Phys.}\ }\textbf {\bibinfo {volume} {101}},\ \bibinfo {pages} {7738}
  (\bibinfo {year} {1994})}\BibitemShut {NoStop}%
\bibitem [{\citenamefont {Klopper}\ and\ \citenamefont
  {Samson}(2002)}]{klopper02}%
  \BibitemOpen
  \bibfield  {author} {\bibinfo {author} {\bibfnamefont {W.}~\bibnamefont
  {Klopper}}\ and\ \bibinfo {author} {\bibfnamefont {C.~C.~M.}\ \bibnamefont
  {Samson}},\ }\href {\doibase 10.1063/1.1461814} {\bibfield  {journal}
  {\bibinfo  {journal} {J. Chem. Phys.}\ }\textbf {\bibinfo {volume} {116}},\
  \bibinfo {pages} {6397} (\bibinfo {year} {2002})}\BibitemShut {NoStop}%
\bibitem [{\citenamefont {Valeev}(2004)}]{valeev04}%
  \BibitemOpen
  \bibfield  {author} {\bibinfo {author} {\bibfnamefont {E.~F.}\ \bibnamefont
  {Valeev}},\ }\href {\doibase 10.1016/j.cplett.2004.07.061} {\bibfield
  {journal} {\bibinfo  {journal} {Chem. Phys. Lett.}\ }\textbf {\bibinfo
  {volume} {395}},\ \bibinfo {pages} {190} (\bibinfo {year}
  {2004})}\BibitemShut {NoStop}%
\bibitem [{\citenamefont {Ten-no}(2004)}]{tenno04}%
  \BibitemOpen
  \bibfield  {author} {\bibinfo {author} {\bibfnamefont {S.}~\bibnamefont
  {Ten-no}},\ }\href {\doibase 10.1016/j.cplett.2004.09.041} {\bibfield
  {journal} {\bibinfo  {journal} {Chem. Phys. Lett.}\ }\textbf {\bibinfo
  {volume} {398}},\ \bibinfo {pages} {56} (\bibinfo {year} {2004})}\BibitemShut
  {NoStop}%
\bibitem [{\citenamefont {Tew}\ \emph {et~al.}(2010)\citenamefont {Tew},
  \citenamefont {H\"attig}, \citenamefont {Bachorz},\ and\ \citenamefont
  {Klopper}}]{tew10_rev}%
  \BibitemOpen
  \bibfield  {author} {\bibinfo {author} {\bibfnamefont {D.~P.}\ \bibnamefont
  {Tew}}, \bibinfo {author} {\bibfnamefont {C.}~\bibnamefont {H\"attig}},
  \bibinfo {author} {\bibfnamefont {R.~A.}\ \bibnamefont {Bachorz}}, \ and\
  \bibinfo {author} {\bibfnamefont {W.}~\bibnamefont {Klopper}},\ }in\
  \href@noop {} {\emph {\bibinfo {booktitle} {Recent Progress in Coupled
  Cluster Methods}}},\ \bibinfo {editor} {edited by\ \bibinfo {editor}
  {\bibfnamefont {P.}~\bibnamefont {\v{C}\'arsky}}, \bibinfo {editor}
  {\bibfnamefont {J.}~\bibnamefont {Paldus}}, \ and\ \bibinfo {editor}
  {\bibfnamefont {J.}~\bibnamefont {Pittner}}}\ (\bibinfo  {publisher}
  {Springer},\ \bibinfo {address} {Dordrecht, Heidelberg, London, New York},\
  \bibinfo {year} {2010})\ pp.\ \bibinfo {pages} {535--572}\BibitemShut
  {NoStop}%
\bibitem [{\citenamefont {K{\"o}hn}, \citenamefont {Richings},\ and\
  \citenamefont {Tew}(2008)}]{koehn08}%
  \BibitemOpen
  \bibfield  {author} {\bibinfo {author} {\bibfnamefont {A.}~\bibnamefont
  {K{\"o}hn}}, \bibinfo {author} {\bibfnamefont {G.~W.}\ \bibnamefont
  {Richings}}, \ and\ \bibinfo {author} {\bibfnamefont {D.~P.}\ \bibnamefont
  {Tew}},\ }\href@noop {} {\bibfield  {journal} {\bibinfo  {journal} {J. Chem.
  Phys.}\ }\textbf {\bibinfo {volume} {129}},\ \bibinfo {pages} {201103}
  (\bibinfo {year} {2008})}\BibitemShut {NoStop}%
\bibitem [{\citenamefont {Shiozaki}\ \emph
  {et~al.}(2008{\natexlab{a}})\citenamefont {Shiozaki}, \citenamefont {Kamiya},
  \citenamefont {Hirata},\ and\ \citenamefont {Valeev}}]{shiozaki08a}%
  \BibitemOpen
  \bibfield  {author} {\bibinfo {author} {\bibfnamefont {T.}~\bibnamefont
  {Shiozaki}}, \bibinfo {author} {\bibfnamefont {M.}~\bibnamefont {Kamiya}},
  \bibinfo {author} {\bibfnamefont {S.}~\bibnamefont {Hirata}}, \ and\ \bibinfo
  {author} {\bibfnamefont {E.~F.}\ \bibnamefont {Valeev}},\ }\href {\doibase
  10.1063/1.2967181} {\bibfield  {journal} {\bibinfo  {journal} {J. Chem.
  Phys.}\ }\textbf {\bibinfo {volume} {129}},\ \bibinfo {pages} {071101}
  (\bibinfo {year} {2008}{\natexlab{a}})}\BibitemShut {NoStop}%
\bibitem [{\citenamefont {Shiozaki}\ \emph
  {et~al.}(2008{\natexlab{b}})\citenamefont {Shiozaki}, \citenamefont {Kamiya},
  \citenamefont {Hirata},\ and\ \citenamefont {Valeev}}]{shiozaki08b}%
  \BibitemOpen
  \bibfield  {author} {\bibinfo {author} {\bibfnamefont {T.}~\bibnamefont
  {Shiozaki}}, \bibinfo {author} {\bibfnamefont {M.}~\bibnamefont {Kamiya}},
  \bibinfo {author} {\bibfnamefont {S.}~\bibnamefont {Hirata}}, \ and\ \bibinfo
  {author} {\bibfnamefont {E.~F.}\ \bibnamefont {Valeev}},\ }\href {\doibase
  10.1039/B803704N} {\bibfield  {journal} {\bibinfo  {journal} {Phys. Chem.
  Chem. Phys.}\ }\textbf {\bibinfo {volume} {10}},\ \bibinfo {pages} {3358}
  (\bibinfo {year} {2008}{\natexlab{b}})}\BibitemShut {NoStop}%
\bibitem [{\citenamefont {Fliegl}, \citenamefont {Klopper},\ and\ \citenamefont
  {H\"attig}(2005)}]{fliegl05}%
  \BibitemOpen
  \bibfield  {author} {\bibinfo {author} {\bibfnamefont {H.}~\bibnamefont
  {Fliegl}}, \bibinfo {author} {\bibfnamefont {W.}~\bibnamefont {Klopper}}, \
  and\ \bibinfo {author} {\bibfnamefont {C.}~\bibnamefont {H\"attig}},\ }\href
  {\doibase 10.1063/1.1850094} {\bibfield  {journal} {\bibinfo  {journal} {J.
  Chem. Phys.}\ }\textbf {\bibinfo {volume} {122}},\ \bibinfo {pages} {084107}
  (\bibinfo {year} {2005})}\BibitemShut {NoStop}%
\bibitem [{\citenamefont {Fliegl}, \citenamefont {H\"attig},\ and\
  \citenamefont {Klopper}(2006)}]{fliegl06}%
  \BibitemOpen
  \bibfield  {author} {\bibinfo {author} {\bibfnamefont {H.}~\bibnamefont
  {Fliegl}}, \bibinfo {author} {\bibfnamefont {C.}~\bibnamefont {H\"attig}}, \
  and\ \bibinfo {author} {\bibfnamefont {W.}~\bibnamefont {Klopper}},\ }\href
  {\doibase 10.1002/qua.20991} {\bibfield  {journal} {\bibinfo  {journal} {Int.
  J. Quantum Chem.}\ }\textbf {\bibinfo {volume} {106}},\ \bibinfo {pages}
  {2306} (\bibinfo {year} {2006})}\BibitemShut {NoStop}%
\bibitem [{\citenamefont {Tew}\ \emph {et~al.}(2007)\citenamefont {Tew},
  \citenamefont {Klopper}, \citenamefont {Neiss},\ and\ \citenamefont
  {H\"attig}}]{tew07}%
  \BibitemOpen
  \bibfield  {author} {\bibinfo {author} {\bibfnamefont {D.~P.}\ \bibnamefont
  {Tew}}, \bibinfo {author} {\bibfnamefont {W.}~\bibnamefont {Klopper}},
  \bibinfo {author} {\bibfnamefont {C.}~\bibnamefont {Neiss}}, \ and\ \bibinfo
  {author} {\bibfnamefont {C.}~\bibnamefont {H\"attig}},\ }\href {\doibase
  10.1039/B617230J} {\bibfield  {journal} {\bibinfo  {journal} {Phys. Chem.
  Chem. Phys.}\ }\textbf {\bibinfo {volume} {9}},\ \bibinfo {pages} {1921}
  (\bibinfo {year} {2007})}\BibitemShut {NoStop}%
\bibitem [{\citenamefont {Tew}\ \emph {et~al.}(2008)\citenamefont {Tew},
  \citenamefont {Klopper}, \citenamefont {Neiss},\ and\ \citenamefont
  {H\"attig}}]{tew08}%
  \BibitemOpen
  \bibfield  {author} {\bibinfo {author} {\bibfnamefont {D.~P.}\ \bibnamefont
  {Tew}}, \bibinfo {author} {\bibfnamefont {W.}~\bibnamefont {Klopper}},
  \bibinfo {author} {\bibfnamefont {C.}~\bibnamefont {Neiss}}, \ and\ \bibinfo
  {author} {\bibfnamefont {C.}~\bibnamefont {H\"attig}},\ }\href {\doibase
  10.1039/b811567b} {\bibfield  {journal} {\bibinfo  {journal} {Phys. Chem.
  Chem. Phys.}\ }\textbf {\bibinfo {volume} {10}},\ \bibinfo {pages} {6325}
  (\bibinfo {year} {2008})}\BibitemShut {NoStop}%
\bibitem [{\citenamefont {Bokhan}, \citenamefont {Bernadotte},\ and\
  \citenamefont {Ten-no}(2009)}]{bokhan09}%
  \BibitemOpen
  \bibfield  {author} {\bibinfo {author} {\bibfnamefont {D.}~\bibnamefont
  {Bokhan}}, \bibinfo {author} {\bibfnamefont {S.}~\bibnamefont {Bernadotte}},
  \ and\ \bibinfo {author} {\bibfnamefont {S.}~\bibnamefont {Ten-no}},\
  }\href@noop {} {\bibfield  {journal} {\bibinfo  {journal} {Chem. Phys.
  Lett.}\ }\textbf {\bibinfo {volume} {469}},\ \bibinfo {pages} {214} (\bibinfo
  {year} {2009})}\BibitemShut {NoStop}%
\bibitem [{\citenamefont {Adler}, \citenamefont {Knizia},\ and\ \citenamefont
  {Werner}(2007)}]{adler07}%
  \BibitemOpen
  \bibfield  {author} {\bibinfo {author} {\bibfnamefont {T.~B.}\ \bibnamefont
  {Adler}}, \bibinfo {author} {\bibfnamefont {G.}~\bibnamefont {Knizia}}, \
  and\ \bibinfo {author} {\bibfnamefont {H.-J.}\ \bibnamefont {Werner}},\
  }\href {\doibase 10.1063/1.2817618} {\bibfield  {journal} {\bibinfo
  {journal} {J. Chem. Phys.}\ }\textbf {\bibinfo {volume} {127}},\ \bibinfo
  {pages} {221106} (\bibinfo {year} {2007})}\BibitemShut {NoStop}%
\bibitem [{\citenamefont {Knizia}, \citenamefont {Adler},\ and\ \citenamefont
  {Werner}(2009)}]{knizia09}%
  \BibitemOpen
  \bibfield  {author} {\bibinfo {author} {\bibfnamefont {G.}~\bibnamefont
  {Knizia}}, \bibinfo {author} {\bibfnamefont {T.~B.}\ \bibnamefont {Adler}}, \
  and\ \bibinfo {author} {\bibfnamefont {H.-J.}\ \bibnamefont {Werner}},\
  }\href {\doibase doi:10.1063/1.3054300} {\bibfield  {journal} {\bibinfo
  {journal} {J. Chem. Phys.}\ }\textbf {\bibinfo {volume} {130}},\ \bibinfo
  {pages} {054104} (\bibinfo {year} {2009})}\BibitemShut {NoStop}%
\bibitem [{\citenamefont {Valeev}(2008)}]{valeev08a}%
  \BibitemOpen
  \bibfield  {author} {\bibinfo {author} {\bibfnamefont {E.~F.}\ \bibnamefont
  {Valeev}},\ }\href {\doibase 10.1039/B713938A} {\bibfield  {journal}
  {\bibinfo  {journal} {Phys. Chem. Chem. Phys.}\ }\textbf {\bibinfo {volume}
  {10}},\ \bibinfo {pages} {106} (\bibinfo {year} {2008})}\BibitemShut
  {NoStop}%
\bibitem [{\citenamefont {Valeev}\ and\ \citenamefont
  {Crawford}(2008)}]{valeev08b}%
  \BibitemOpen
  \bibfield  {author} {\bibinfo {author} {\bibfnamefont {E.~F.}\ \bibnamefont
  {Valeev}}\ and\ \bibinfo {author} {\bibfnamefont {T.~D.}\ \bibnamefont
  {Crawford}},\ }\href {\doibase 10.1063/1.2939577} {\bibfield  {journal}
  {\bibinfo  {journal} {J. Chem. Phys.}\ }\textbf {\bibinfo {volume} {128}},\
  \bibinfo {pages} {244113} (\bibinfo {year} {2008})}\BibitemShut {NoStop}%
\bibitem [{\citenamefont {Torheyden}\ and\ \citenamefont
  {Valeev}(2009)}]{torheyden09}%
  \BibitemOpen
  \bibfield  {author} {\bibinfo {author} {\bibfnamefont {M.}~\bibnamefont
  {Torheyden}}\ and\ \bibinfo {author} {\bibfnamefont {E.~F.}\ \bibnamefont
  {Valeev}},\ }\href@noop {} {\bibfield  {journal} {\bibinfo  {journal} {J.
  Chem. Phys.}\ }\textbf {\bibinfo {volume} {131}},\ \bibinfo {pages} {171103}
  (\bibinfo {year} {2009})}\BibitemShut {NoStop}%
\bibitem [{\citenamefont {H\"attig}, \citenamefont {Tew},\ and\ \citenamefont
  {K\"ohn}(2010)}]{haettig10}%
  \BibitemOpen
  \bibfield  {author} {\bibinfo {author} {\bibfnamefont {C.}~\bibnamefont
  {H\"attig}}, \bibinfo {author} {\bibfnamefont {D.~P.}\ \bibnamefont {Tew}}, \
  and\ \bibinfo {author} {\bibfnamefont {A.}~\bibnamefont {K\"ohn}},\
  }\href@noop {} {\bibfield  {journal} {\bibinfo  {journal} {J. Chem. Phys.}\
  }\textbf {\bibinfo {volume} {132}},\ \bibinfo {pages} {231102} (\bibinfo
  {year} {2010})}\BibitemShut {NoStop}%
\bibitem [{\citenamefont {Werner}\ \emph {et~al.}(2010)\citenamefont {Werner},
  \citenamefont {Adler}, \citenamefont {Knizia},\ and\ \citenamefont
  {Manby}}]{werner10_rev}%
  \BibitemOpen
  \bibfield  {author} {\bibinfo {author} {\bibfnamefont {H.-J.}\ \bibnamefont
  {Werner}}, \bibinfo {author} {\bibfnamefont {T.~B.}\ \bibnamefont {Adler}},
  \bibinfo {author} {\bibfnamefont {G.}~\bibnamefont {Knizia}}, \ and\ \bibinfo
  {author} {\bibfnamefont {F.~R.}\ \bibnamefont {Manby}},\ }in\ \href@noop {}
  {\emph {\bibinfo {booktitle} {Recent Progress in Coupled Cluster Methods}}},\
  \bibinfo {editor} {edited by\ \bibinfo {editor} {\bibfnamefont
  {P.}~\bibnamefont {\v{C}\'arsky}}, \bibinfo {editor} {\bibfnamefont
  {J.}~\bibnamefont {Paldus}}, \ and\ \bibinfo {editor} {\bibfnamefont
  {J.}~\bibnamefont {Pittner}}}\ (\bibinfo  {publisher} {Springer},\ \bibinfo
  {address} {Dordrecht, Heidelberg, London, New York},\ \bibinfo {year}
  {2010})\ pp.\ \bibinfo {pages} {573--619}\BibitemShut {NoStop}%
\bibitem [{\citenamefont {K\"ohn}(2009)}]{koehn09}%
  \BibitemOpen
  \bibfield  {author} {\bibinfo {author} {\bibfnamefont {A.}~\bibnamefont
  {K\"ohn}},\ }\href {\doibase 10.1063/1.3116792} {\bibfield  {journal}
  {\bibinfo  {journal} {J. Chem. Phys.}\ }\textbf {\bibinfo {volume} {130}},\
  \bibinfo {pages} {131101} (\bibinfo {year} {2009})}\BibitemShut {NoStop}%
\bibitem [{\citenamefont {K{\"o}hn}(2010)}]{koehn10}%
  \BibitemOpen
  \bibfield  {author} {\bibinfo {author} {\bibfnamefont {A.}~\bibnamefont
  {K{\"o}hn}},\ }\href {\doibase 10.1063/1.3496373} {\bibfield  {journal}
  {\bibinfo  {journal} {J. Chem. Phys.}\ }\textbf {\bibinfo {volume} {133}},\
  \bibinfo {pages} {174118} (\bibinfo {year} {2010})}\BibitemShut {NoStop}%
\bibitem [{\citenamefont {Boys}\ and\ \citenamefont
  {Handy}(1969{\natexlab{a}})}]{boys69}%
  \BibitemOpen
  \bibfield  {author} {\bibinfo {author} {\bibfnamefont {S.~F.}\ \bibnamefont
  {Boys}}\ and\ \bibinfo {author} {\bibfnamefont {N.~C.}\ \bibnamefont
  {Handy}},\ }\href {\doibase 10.1098/rspa.1969.0062} {\bibfield  {journal}
  {\bibinfo  {journal} {Proc. R. Soc. A}\ }\textbf {\bibinfo {volume} {310}},\
  \bibinfo {pages} {63} (\bibinfo {year} {1969}{\natexlab{a}})}\BibitemShut
  {NoStop}%
\bibitem [{\citenamefont {Hirschfelder}(1963)}]{hirschfelder63}%
  \BibitemOpen
  \bibfield  {author} {\bibinfo {author} {\bibfnamefont {J.~O.}\ \bibnamefont
  {Hirschfelder}},\ }\href {\doibase 10.1063/1.1734157} {\bibfield  {journal}
  {\bibinfo  {journal} {J. Chem. Phys.}\ }\textbf {\bibinfo {volume} {39}},\
  \bibinfo {pages} {3145} (\bibinfo {year} {1963})}\BibitemShut {NoStop}%
\bibitem [{\citenamefont {Boys}\ and\ \citenamefont
  {Handy}(1969{\natexlab{b}})}]{boys69_full}%
  \BibitemOpen
  \bibfield  {author} {\bibinfo {author} {\bibfnamefont {S.~F.}\ \bibnamefont
  {Boys}}\ and\ \bibinfo {author} {\bibfnamefont {N.~C.}\ \bibnamefont
  {Handy}},\ }\href {\doibase 10.1098/rspa.1969.0061} {\bibfield  {journal}
  {\bibinfo  {journal} {Proc. R. Soc. A}\ }\textbf {\bibinfo {volume} {310}},\
  \bibinfo {pages} {43} (\bibinfo {year} {1969}{\natexlab{b}})}\BibitemShut
  {NoStop}%
\bibitem [{\citenamefont {Boys}\ and\ \citenamefont
  {Handy}(1969{\natexlab{c}})}]{boys69_indeterminacy}%
  \BibitemOpen
  \bibfield  {author} {\bibinfo {author} {\bibfnamefont {S.~F.}\ \bibnamefont
  {Boys}}\ and\ \bibinfo {author} {\bibfnamefont {N.~C.}\ \bibnamefont
  {Handy}},\ }\href {\doibase 10.1098/rspa.1969.0038} {\bibfield  {journal}
  {\bibinfo  {journal} {Proc. R. Soc. A}\ }\textbf {\bibinfo {volume} {309}},\
  \bibinfo {pages} {209} (\bibinfo {year} {1969}{\natexlab{c}})}\BibitemShut
  {NoStop}%
\bibitem [{\citenamefont {Boys}(1969)}]{boys69_bilinear}%
  \BibitemOpen
  \bibfield  {author} {\bibinfo {author} {\bibfnamefont {S.~F.}\ \bibnamefont
  {Boys}},\ }\href {\doibase 10.1098/rspa.1969.0037} {\bibfield  {journal}
  {\bibinfo  {journal} {Proc. R. Soc. A}\ }\textbf {\bibinfo {volume} {309}},\
  \bibinfo {pages} {195} (\bibinfo {year} {1969})}\BibitemShut {NoStop}%
\bibitem [{\citenamefont {Boys}\ and\ \citenamefont
  {Handy}(1969{\natexlab{d}})}]{boys69_lih}%
  \BibitemOpen
  \bibfield  {author} {\bibinfo {author} {\bibfnamefont {S.~F.}\ \bibnamefont
  {Boys}}\ and\ \bibinfo {author} {\bibfnamefont {N.~C.}\ \bibnamefont
  {Handy}},\ }\href {\doibase 10.1098/rspa.1969.0120} {\bibfield  {journal}
  {\bibinfo  {journal} {Proc. R. Soc. A}\ }\textbf {\bibinfo {volume} {311}},\
  \bibinfo {pages} {309} (\bibinfo {year} {1969}{\natexlab{d}})}\BibitemShut
  {NoStop}%
\bibitem [{\citenamefont {Handy}(1969)}]{handy69}%
  \BibitemOpen
  \bibfield  {author} {\bibinfo {author} {\bibfnamefont {N.~C.}\ \bibnamefont
  {Handy}},\ }\href {\doibase 10.1063/1.1672496} {\bibfield  {journal}
  {\bibinfo  {journal} {J. Chem. Phys.}\ }\textbf {\bibinfo {volume} {51}},\
  \bibinfo {pages} {3205} (\bibinfo {year} {1969})}\BibitemShut {NoStop}%
\bibitem [{\citenamefont {Handy}(1971)}]{handy71}%
  \BibitemOpen
  \bibfield  {author} {\bibinfo {author} {\bibfnamefont {N.~C.}\ \bibnamefont
  {Handy}},\ }\href {\doibase 10.1080/00268977100101961} {\bibfield  {journal}
  {\bibinfo  {journal} {Mol. Phys.}\ }\textbf {\bibinfo {volume} {21}},\
  \bibinfo {pages} {817} (\bibinfo {year} {1971})}\BibitemShut {NoStop}%
\bibitem [{\citenamefont {Handy}(1972)}]{handy72}%
  \BibitemOpen
  \bibfield  {author} {\bibinfo {author} {\bibfnamefont {N.~C.}\ \bibnamefont
  {Handy}},\ }\href {\doibase 10.1080/00268977200100011} {\bibfield  {journal}
  {\bibinfo  {journal} {Mol. Phys.}\ }\textbf {\bibinfo {volume} {23}},\
  \bibinfo {pages} {1} (\bibinfo {year} {1972})}\BibitemShut {NoStop}%
\bibitem [{\citenamefont {Handy}(1973)}]{handy73}%
  \BibitemOpen
  \bibfield  {author} {\bibinfo {author} {\bibfnamefont {N.~C.}\ \bibnamefont
  {Handy}},\ }\href {\doibase 10.1063/1.1678918} {\bibfield  {journal}
  {\bibinfo  {journal} {J. Chem. Phys.}\ }\textbf {\bibinfo {volume} {58}},\
  \bibinfo {pages} {279} (\bibinfo {year} {1973})}\BibitemShut {NoStop}%
\bibitem [{\citenamefont {Nooijen}\ and\ \citenamefont
  {Bartlett}(1998)}]{nooijen98}%
  \BibitemOpen
  \bibfield  {author} {\bibinfo {author} {\bibfnamefont {M.}~\bibnamefont
  {Nooijen}}\ and\ \bibinfo {author} {\bibfnamefont {R.~J.}\ \bibnamefont
  {Bartlett}},\ }\href {\doibase 10.1063/1.477485} {\bibfield  {journal}
  {\bibinfo  {journal} {J. Chem. Phys.}\ }\textbf {\bibinfo {volume} {109}},\
  \bibinfo {pages} {8232} (\bibinfo {year} {1998})}\BibitemShut {NoStop}%
\bibitem [{\citenamefont {{Ten-no}}(2000)}]{ten-no00}%
  \BibitemOpen
  \bibfield  {author} {\bibinfo {author} {\bibfnamefont {S.}~\bibnamefont
  {{Ten-no}}},\ }\href {\doibase 10.1016/S0009-2614(00)01066-6} {\bibfield
  {journal} {\bibinfo  {journal} {Chem. Phys. Lett.}\ }\textbf {\bibinfo
  {volume} {330}},\ \bibinfo {pages} {169} (\bibinfo {year}
  {2000})}\BibitemShut {NoStop}%
\bibitem [{\citenamefont {Ten-no}(2000)}]{ten-no00_integrals}%
  \BibitemOpen
  \bibfield  {author} {\bibinfo {author} {\bibfnamefont {S.}~\bibnamefont
  {Ten-no}},\ }\href {\doibase 10.1016/S0009-2614(00)01067-8} {\bibfield
  {journal} {\bibinfo  {journal} {Chem. Phys. Lett.}\ }\textbf {\bibinfo
  {volume} {330}},\ \bibinfo {pages} {175} (\bibinfo {year}
  {2000})}\BibitemShut {NoStop}%
\bibitem [{\citenamefont {Hino}, \citenamefont {Tanimura},\ and\ \citenamefont
  {{Ten-no}}(2002)}]{hino02}%
  \BibitemOpen
  \bibfield  {author} {\bibinfo {author} {\bibfnamefont {O.}~\bibnamefont
  {Hino}}, \bibinfo {author} {\bibfnamefont {Y.}~\bibnamefont {Tanimura}}, \
  and\ \bibinfo {author} {\bibfnamefont {S.}~\bibnamefont {{Ten-no}}},\ }\href
  {\doibase 10.1016/S0009-2614(02)00042-8} {\bibfield  {journal} {\bibinfo
  {journal} {Chem. Phys. Lett.}\ }\textbf {\bibinfo {volume} {353}},\ \bibinfo
  {pages} {317} (\bibinfo {year} {2002})}\BibitemShut {NoStop}%
\bibitem [{\citenamefont {Imamura}\ and\ \citenamefont
  {Scuseria}(2003)}]{imamura03}%
  \BibitemOpen
  \bibfield  {author} {\bibinfo {author} {\bibfnamefont {Y.}~\bibnamefont
  {Imamura}}\ and\ \bibinfo {author} {\bibfnamefont {G.~E.}\ \bibnamefont
  {Scuseria}},\ }\href {\doibase 10.1063/1.1535442} {\bibfield  {journal}
  {\bibinfo  {journal} {J. Chem. Phys.}\ }\textbf {\bibinfo {volume} {118}},\
  \bibinfo {pages} {2464} (\bibinfo {year} {2003})}\BibitemShut {NoStop}%
\bibitem [{\citenamefont {Zweistra}, \citenamefont {Samson},\ and\
  \citenamefont {Klopper}(2003)}]{zweistra03}%
  \BibitemOpen
  \bibfield  {author} {\bibinfo {author} {\bibfnamefont {H.~J.~A.}\
  \bibnamefont {Zweistra}}, \bibinfo {author} {\bibfnamefont {C.~C.~M.}\
  \bibnamefont {Samson}}, \ and\ \bibinfo {author} {\bibfnamefont
  {W.}~\bibnamefont {Klopper}},\ }\href {\doibase 10.1135/cccc20030374}
  {\bibfield  {journal} {\bibinfo  {journal} {Collect. Czech. Chem. Commun.}\
  }\textbf {\bibinfo {volume} {68}},\ \bibinfo {pages} {374} (\bibinfo {year}
  {2003})}\BibitemShut {NoStop}%
\bibitem [{\citenamefont {Umezawa}\ and\ \citenamefont
  {Tsuneyuki}(2003)}]{umezawa03}%
  \BibitemOpen
  \bibfield  {author} {\bibinfo {author} {\bibfnamefont {N.}~\bibnamefont
  {Umezawa}}\ and\ \bibinfo {author} {\bibfnamefont {S.}~\bibnamefont
  {Tsuneyuki}},\ }\href {\doibase 10.1063/1.1617274} {\bibfield  {journal}
  {\bibinfo  {journal} {J. Chem. Phys.}\ }\textbf {\bibinfo {volume} {119}},\
  \bibinfo {pages} {10015} (\bibinfo {year} {2003})}\BibitemShut {NoStop}%
\bibitem [{\citenamefont {Umezawa}\ and\ \citenamefont
  {Tsuneyuki}(2004{\natexlab{a}})}]{umezawa04_excited}%
  \BibitemOpen
  \bibfield  {author} {\bibinfo {author} {\bibfnamefont {N.}~\bibnamefont
  {Umezawa}}\ and\ \bibinfo {author} {\bibfnamefont {S.}~\bibnamefont
  {Tsuneyuki}},\ }\href {\doibase 10.1063/1.1792392} {\bibfield  {journal}
  {\bibinfo  {journal} {J. Chem. Phys.}\ }\textbf {\bibinfo {volume} {121}},\
  \bibinfo {pages} {7070} (\bibinfo {year} {2004}{\natexlab{a}})}\BibitemShut
  {NoStop}%
\bibitem [{\citenamefont {Umezawa}\ and\ \citenamefont
  {Tsuneyuki}(2004{\natexlab{b}})}]{umezawa04_ueg}%
  \BibitemOpen
  \bibfield  {author} {\bibinfo {author} {\bibfnamefont {N.}~\bibnamefont
  {Umezawa}}\ and\ \bibinfo {author} {\bibfnamefont {S.}~\bibnamefont
  {Tsuneyuki}},\ }\href {\doibase 10.1103/PhysRevB.69.165102} {\bibfield
  {journal} {\bibinfo  {journal} {Phys. Rev. B}\ }\textbf {\bibinfo {volume}
  {69}},\ \bibinfo {pages} {165102} (\bibinfo {year}
  {2004}{\natexlab{b}})}\BibitemShut {NoStop}%
\bibitem [{\citenamefont {Umezawa}\ \emph {et~al.}(2005)\citenamefont
  {Umezawa}, \citenamefont {Tsuneyuki}, \citenamefont {Ohno}, \citenamefont
  {Shiraishi},\ and\ \citenamefont {Chikyow}}]{umezawa05}%
  \BibitemOpen
  \bibfield  {author} {\bibinfo {author} {\bibfnamefont {N.}~\bibnamefont
  {Umezawa}}, \bibinfo {author} {\bibfnamefont {S.}~\bibnamefont {Tsuneyuki}},
  \bibinfo {author} {\bibfnamefont {T.}~\bibnamefont {Ohno}}, \bibinfo {author}
  {\bibfnamefont {K.}~\bibnamefont {Shiraishi}}, \ and\ \bibinfo {author}
  {\bibfnamefont {T.}~\bibnamefont {Chikyow}},\ }\href {\doibase
  10.1063/1.1924597} {\bibfield  {journal} {\bibinfo  {journal} {J. Chem.
  Phys.}\ }\textbf {\bibinfo {volume} {122}},\ \bibinfo {pages} {224101}
  (\bibinfo {year} {2005})}\BibitemShut {NoStop}%
\bibitem [{\citenamefont {Yanai}\ and\ \citenamefont {Chan}(2006)}]{yanai06}%
  \BibitemOpen
  \bibfield  {author} {\bibinfo {author} {\bibfnamefont {T.}~\bibnamefont
  {Yanai}}\ and\ \bibinfo {author} {\bibfnamefont {G.~K.-L.}\ \bibnamefont
  {Chan}},\ }\href {\doibase 10.1063/1.2196410} {\bibfield  {journal} {\bibinfo
   {journal} {J. Chem. Phys.}\ }\textbf {\bibinfo {volume} {124}},\ \bibinfo
  {pages} {194106} (\bibinfo {year} {2006})}\BibitemShut {NoStop}%
\bibitem [{\citenamefont {Yanai}\ and\ \citenamefont {Chan}(2007)}]{yanai07}%
  \BibitemOpen
  \bibfield  {author} {\bibinfo {author} {\bibfnamefont {T.}~\bibnamefont
  {Yanai}}\ and\ \bibinfo {author} {\bibfnamefont {G.~K.-L.}\ \bibnamefont
  {Chan}},\ }\href {\doibase 10.1063/1.2761870} {\bibfield  {journal} {\bibinfo
   {journal} {J. Chem. Phys.}\ }\textbf {\bibinfo {volume} {127}},\ \bibinfo
  {pages} {104107} (\bibinfo {year} {2007})}\BibitemShut {NoStop}%
\bibitem [{\citenamefont {Neuscamman}, \citenamefont {Yanai},\ and\
  \citenamefont {Chan}(2010)}]{neuscamman10}%
  \BibitemOpen
  \bibfield  {author} {\bibinfo {author} {\bibfnamefont {E.}~\bibnamefont
  {Neuscamman}}, \bibinfo {author} {\bibfnamefont {T.}~\bibnamefont {Yanai}}, \
  and\ \bibinfo {author} {\bibfnamefont {G.~K.-L.}\ \bibnamefont {Chan}},\
  }\href {\doibase 10.1080/01442351003620540} {\bibfield  {journal} {\bibinfo
  {journal} {Int. Rev. Phys. Chem.}\ }\textbf {\bibinfo {volume} {29}},\
  \bibinfo {pages} {231} (\bibinfo {year} {2010})}\BibitemShut {NoStop}%
\bibitem [{\citenamefont {Tsuneyuki}(2008)}]{tsuneyuki08}%
  \BibitemOpen
  \bibfield  {author} {\bibinfo {author} {\bibfnamefont {S.}~\bibnamefont
  {Tsuneyuki}},\ }\href {\doibase 10.1143/PTPS.176.134} {\bibfield  {journal}
  {\bibinfo  {journal} {Prog. Theor. Phys. Suppl.}\ }\textbf {\bibinfo {volume}
  {176}},\ \bibinfo {pages} {134} (\bibinfo {year} {2008})}\BibitemShut
  {NoStop}%
\bibitem [{\citenamefont {Luo}(2010)}]{luo10}%
  \BibitemOpen
  \bibfield  {author} {\bibinfo {author} {\bibfnamefont {H.}~\bibnamefont
  {Luo}},\ }\href {\doibase 10.1063/1.3505037} {\bibfield  {journal} {\bibinfo
  {journal} {J. Chem. Phys.}\ }\textbf {\bibinfo {volume} {133}},\ \bibinfo
  {pages} {154109} (\bibinfo {year} {2010})}\BibitemShut {NoStop}%
\bibitem [{\citenamefont {Luo}, \citenamefont {Hackbusch},\ and\ \citenamefont
  {Flad}(2010)}]{luo10_study}%
  \BibitemOpen
  \bibfield  {author} {\bibinfo {author} {\bibfnamefont {H.}~\bibnamefont
  {Luo}}, \bibinfo {author} {\bibfnamefont {W.}~\bibnamefont {Hackbusch}}, \
  and\ \bibinfo {author} {\bibfnamefont {H.-J.}\ \bibnamefont {Flad}},\ }\href
  {\doibase 10.1080/00268970903521194} {\bibfield  {journal} {\bibinfo
  {journal} {Mol. Phys.}\ }\textbf {\bibinfo {volume} {108}},\ \bibinfo {pages}
  {425} (\bibinfo {year} {2010})}\BibitemShut {NoStop}%
\bibitem [{\citenamefont {Luo}(2011)}]{luo11}%
  \BibitemOpen
  \bibfield  {author} {\bibinfo {author} {\bibfnamefont {H.}~\bibnamefont
  {Luo}},\ }\href {\doibase 10.1063/1.3607990} {\bibfield  {journal} {\bibinfo
  {journal} {J. Chem. Phys.}\ }\textbf {\bibinfo {volume} {135}},\ \bibinfo
  {pages} {024109} (\bibinfo {year} {2011})}\BibitemShut {NoStop}%
\bibitem [{\citenamefont {Luo}(2012)}]{luo12}%
  \BibitemOpen
  \bibfield  {author} {\bibinfo {author} {\bibfnamefont {H.}~\bibnamefont
  {Luo}},\ }\href {\doibase 10.1063/1.4727852} {\bibfield  {journal} {\bibinfo
  {journal} {J. Chem. Phys.}\ }\textbf {\bibinfo {volume} {136}},\ \bibinfo
  {pages} {224111} (\bibinfo {year} {2012})}\BibitemShut {NoStop}%
\bibitem [{\citenamefont {Yanai}\ and\ \citenamefont
  {Shiozaki}(2012)}]{yanai12}%
  \BibitemOpen
  \bibfield  {author} {\bibinfo {author} {\bibfnamefont {T.}~\bibnamefont
  {Yanai}}\ and\ \bibinfo {author} {\bibfnamefont {T.}~\bibnamefont
  {Shiozaki}},\ }\href {\doibase 10.1063/1.3688225} {\bibfield  {journal}
  {\bibinfo  {journal} {J. Chem. Phys.}\ }\textbf {\bibinfo {volume} {136}},\
  \bibinfo {pages} {084107} (\bibinfo {year} {2012})}\BibitemShut {NoStop}%
\bibitem [{\citenamefont {Sharma}\ \emph {et~al.}(2014)\citenamefont {Sharma},
  \citenamefont {Yanai}, \citenamefont {Booth}, \citenamefont {Umrigar},\ and\
  \citenamefont {Chan}}]{sharma14}%
  \BibitemOpen
  \bibfield  {author} {\bibinfo {author} {\bibfnamefont {S.}~\bibnamefont
  {Sharma}}, \bibinfo {author} {\bibfnamefont {T.}~\bibnamefont {Yanai}},
  \bibinfo {author} {\bibfnamefont {G.~H.}\ \bibnamefont {Booth}}, \bibinfo
  {author} {\bibfnamefont {C.~J.}\ \bibnamefont {Umrigar}}, \ and\ \bibinfo
  {author} {\bibfnamefont {G.~K.-L.}\ \bibnamefont {Chan}},\ }\href {\doibase
  10.1063/1.4867383} {\bibfield  {journal} {\bibinfo  {journal} {J. Chem.
  Phys.}\ }\textbf {\bibinfo {volume} {140}},\ \bibinfo {pages} {104112}
  (\bibinfo {year} {2014})}\BibitemShut {NoStop}%
\bibitem [{\citenamefont {Ochi}\ \emph {et~al.}(2012)\citenamefont {Ochi},
  \citenamefont {Sodeyama}, \citenamefont {Sakuma},\ and\ \citenamefont
  {Tsuneyuki}}]{ochi12}%
  \BibitemOpen
  \bibfield  {author} {\bibinfo {author} {\bibfnamefont {M.}~\bibnamefont
  {Ochi}}, \bibinfo {author} {\bibfnamefont {K.}~\bibnamefont {Sodeyama}},
  \bibinfo {author} {\bibfnamefont {R.}~\bibnamefont {Sakuma}}, \ and\ \bibinfo
  {author} {\bibfnamefont {S.}~\bibnamefont {Tsuneyuki}},\ }\href {\doibase
  10.1063/1.3689440} {\bibfield  {journal} {\bibinfo  {journal} {J. Chem.
  Phys.}\ }\textbf {\bibinfo {volume} {136}},\ \bibinfo {pages} {094108}
  (\bibinfo {year} {2012})}\BibitemShut {NoStop}%
\bibitem [{\citenamefont {Ochi}\ and\ \citenamefont
  {Tsuneyuki}(2014)}]{ochi14}%
  \BibitemOpen
  \bibfield  {author} {\bibinfo {author} {\bibfnamefont {M.}~\bibnamefont
  {Ochi}}\ and\ \bibinfo {author} {\bibfnamefont {S.}~\bibnamefont
  {Tsuneyuki}},\ }\href {\doibase 10.1021/ct500485b} {\bibfield  {journal}
  {\bibinfo  {journal} {J. Chem. Theory Comput.}\ }\textbf {\bibinfo {volume}
  {10}},\ \bibinfo {pages} {4098} (\bibinfo {year} {2014})}\BibitemShut
  {NoStop}%
\bibitem [{\citenamefont {Ochi}\ and\ \citenamefont
  {Tsuneyuki}(2015)}]{ochi15}%
  \BibitemOpen
  \bibfield  {author} {\bibinfo {author} {\bibfnamefont {M.}~\bibnamefont
  {Ochi}}\ and\ \bibinfo {author} {\bibfnamefont {S.}~\bibnamefont
  {Tsuneyuki}},\ }\href {\doibase 10.1016/j.cplett.2015.01.009} {\bibfield
  {journal} {\bibinfo  {journal} {Chem. Phys. Lett.}\ }\textbf {\bibinfo
  {volume} {621}},\ \bibinfo {pages} {177} (\bibinfo {year}
  {2015})}\BibitemShut {NoStop}%
\bibitem [{\citenamefont {Yanai}\ \emph {et~al.}(2015)\citenamefont {Yanai},
  \citenamefont {Kurashige}, \citenamefont {Mizukami}, \citenamefont
  {Chalupský}, \citenamefont {Lan},\ and\ \citenamefont {Saitow}}]{yanai15}%
  \BibitemOpen
  \bibfield  {author} {\bibinfo {author} {\bibfnamefont {T.}~\bibnamefont
  {Yanai}}, \bibinfo {author} {\bibfnamefont {Y.}~\bibnamefont {Kurashige}},
  \bibinfo {author} {\bibfnamefont {W.}~\bibnamefont {Mizukami}}, \bibinfo
  {author} {\bibfnamefont {J.}~\bibnamefont {Chalupský}}, \bibinfo {author}
  {\bibfnamefont {T.~N.}\ \bibnamefont {Lan}}, \ and\ \bibinfo {author}
  {\bibfnamefont {M.}~\bibnamefont {Saitow}},\ }\href {\doibase
  10.1002/qua.24808} {\bibfield  {journal} {\bibinfo  {journal} {Int. J.
  Quantum Chem.}\ }\textbf {\bibinfo {volume} {115}},\ \bibinfo {pages} {283}
  (\bibinfo {year} {2015})}\BibitemShut {NoStop}%
\bibitem [{\citenamefont {Ochi}\ \emph {et~al.}(2016)\citenamefont {Ochi},
  \citenamefont {Yamamoto}, \citenamefont {Arita},\ and\ \citenamefont
  {Tsuneyuki}}]{ochi16}%
  \BibitemOpen
  \bibfield  {author} {\bibinfo {author} {\bibfnamefont {M.}~\bibnamefont
  {Ochi}}, \bibinfo {author} {\bibfnamefont {Y.}~\bibnamefont {Yamamoto}},
  \bibinfo {author} {\bibfnamefont {R.}~\bibnamefont {Arita}}, \ and\ \bibinfo
  {author} {\bibfnamefont {S.}~\bibnamefont {Tsuneyuki}},\ }\href {\doibase
  10.1063/1.4943117} {\bibfield  {journal} {\bibinfo  {journal} {J. Chem.
  Phys.}\ }\textbf {\bibinfo {volume} {144}},\ \bibinfo {pages} {104109}
  (\bibinfo {year} {2016})}\BibitemShut {NoStop}%
\bibitem [{\citenamefont {Wahlen-Strothman}\ \emph {et~al.}(2015)\citenamefont
  {Wahlen-Strothman}, \citenamefont {Jim{\'e}nez-Hoyos}, \citenamefont
  {Henderson},\ and\ \citenamefont {Scuseria}}]{wahlen-strothman15}%
  \BibitemOpen
  \bibfield  {author} {\bibinfo {author} {\bibfnamefont {J.~M.}\ \bibnamefont
  {Wahlen-Strothman}}, \bibinfo {author} {\bibfnamefont {C.~A.}\ \bibnamefont
  {Jim{\'e}nez-Hoyos}}, \bibinfo {author} {\bibfnamefont {T.~M.}\ \bibnamefont
  {Henderson}}, \ and\ \bibinfo {author} {\bibfnamefont {G.~E.}\ \bibnamefont
  {Scuseria}},\ }\href {\doibase 10.1103/PhysRevB.91.041114} {\bibfield
  {journal} {\bibinfo  {journal} {Phys. Rev. B}\ }\textbf {\bibinfo {volume}
  {91}},\ \bibinfo {pages} {041114} (\bibinfo {year} {2015})}\BibitemShut
  {NoStop}%
\bibitem [{\citenamefont {Kersten}, \citenamefont {Booth},\ and\ \citenamefont
  {Alavi}(2016)}]{kersten16}%
  \BibitemOpen
  \bibfield  {author} {\bibinfo {author} {\bibfnamefont {J.~A.~F.}\
  \bibnamefont {Kersten}}, \bibinfo {author} {\bibfnamefont {G.~H.}\
  \bibnamefont {Booth}}, \ and\ \bibinfo {author} {\bibfnamefont
  {A.}~\bibnamefont {Alavi}},\ }\href {\doibase 10.1063/1.4959245} {\bibfield
  {journal} {\bibinfo  {journal} {J. Chem. Phys.}\ }\textbf {\bibinfo {volume}
  {145}},\ \bibinfo {pages} {054117} (\bibinfo {year} {2016})}\BibitemShut
  {NoStop}%
\bibitem [{\citenamefont {Jeszenszki}\ \emph {et~al.}(2018)\citenamefont
  {Jeszenszki}, \citenamefont {Luo}, \citenamefont {Alavi},\ and\ \citenamefont
  {Brand}}]{jeszenszki18}%
  \BibitemOpen
  \bibfield  {author} {\bibinfo {author} {\bibfnamefont {P.}~\bibnamefont
  {Jeszenszki}}, \bibinfo {author} {\bibfnamefont {H.}~\bibnamefont {Luo}},
  \bibinfo {author} {\bibfnamefont {A.}~\bibnamefont {Alavi}}, \ and\ \bibinfo
  {author} {\bibfnamefont {J.}~\bibnamefont {Brand}},\ }\href {\doibase
  10.1103/PhysRevA.98.053627} {\bibfield  {journal} {\bibinfo  {journal} {Phys.
  Rev. A}\ }\textbf {\bibinfo {volume} {98}},\ \bibinfo {pages} {053627}
  (\bibinfo {year} {2018})}\BibitemShut {NoStop}%
\bibitem [{\citenamefont {Luo}\ and\ \citenamefont {Alavi}(2018)}]{luo18}%
  \BibitemOpen
  \bibfield  {author} {\bibinfo {author} {\bibfnamefont {H.}~\bibnamefont
  {Luo}}\ and\ \bibinfo {author} {\bibfnamefont {A.}~\bibnamefont {Alavi}},\
  }\href {\doibase 10.1021/acs.jctc.7b01257} {\bibfield  {journal} {\bibinfo
  {journal} {J. Chem. Theory Comput.}\ }\textbf {\bibinfo {volume} {14}},\
  \bibinfo {pages} {1403} (\bibinfo {year} {2018})}\BibitemShut {NoStop}%
\bibitem [{\citenamefont {Dobrautz}, \citenamefont {Luo},\ and\ \citenamefont
  {Alavi}(2019)}]{dobrautz19}%
  \BibitemOpen
  \bibfield  {author} {\bibinfo {author} {\bibfnamefont {W.}~\bibnamefont
  {Dobrautz}}, \bibinfo {author} {\bibfnamefont {H.}~\bibnamefont {Luo}}, \
  and\ \bibinfo {author} {\bibfnamefont {A.}~\bibnamefont {Alavi}},\ }\href
  {\doibase 10.1103/PhysRevB.99.075119} {\bibfield  {journal} {\bibinfo
  {journal} {Phys. Rev. B}\ }\textbf {\bibinfo {volume} {99}},\ \bibinfo
  {pages} {075119} (\bibinfo {year} {2019})}\BibitemShut {NoStop}%
\bibitem [{\citenamefont {Cohen}\ \emph {et~al.}(2019)\citenamefont {Cohen},
  \citenamefont {Luo}, \citenamefont {Guther}, \citenamefont {Dobrautz},
  \citenamefont {Tew},\ and\ \citenamefont {Alavi}}]{cohen19}%
  \BibitemOpen
  \bibfield  {author} {\bibinfo {author} {\bibfnamefont {A.~J.}\ \bibnamefont
  {Cohen}}, \bibinfo {author} {\bibfnamefont {H.}~\bibnamefont {Luo}}, \bibinfo
  {author} {\bibfnamefont {K.}~\bibnamefont {Guther}}, \bibinfo {author}
  {\bibfnamefont {W.}~\bibnamefont {Dobrautz}}, \bibinfo {author}
  {\bibfnamefont {D.~P.}\ \bibnamefont {Tew}}, \ and\ \bibinfo {author}
  {\bibfnamefont {A.}~\bibnamefont {Alavi}},\ }\href {\doibase
  10.1063/1.5116024} {\bibfield  {journal} {\bibinfo  {journal} {J. Chem.
  Phys.}\ }\textbf {\bibinfo {volume} {151}},\ \bibinfo {pages} {061101}
  (\bibinfo {year} {2019})}\BibitemShut {NoStop}%
\bibitem [{\citenamefont {Guther}\ \emph {et~al.}(2021)\citenamefont {Guther},
  \citenamefont {Cohen}, \citenamefont {Luo},\ and\ \citenamefont
  {Alavi}}]{guther21}%
  \BibitemOpen
  \bibfield  {author} {\bibinfo {author} {\bibfnamefont {K.}~\bibnamefont
  {Guther}}, \bibinfo {author} {\bibfnamefont {A.~J.}\ \bibnamefont {Cohen}},
  \bibinfo {author} {\bibfnamefont {H.}~\bibnamefont {Luo}}, \ and\ \bibinfo
  {author} {\bibfnamefont {A.}~\bibnamefont {Alavi}},\ }\href {\doibase
  10.1063/5.0055575} {\bibfield  {journal} {\bibinfo  {journal} {J. Chem.
  Phys.}\ }\textbf {\bibinfo {volume} {155}},\ \bibinfo {pages} {011102}
  (\bibinfo {year} {2021})}\BibitemShut {NoStop}%
\bibitem [{\citenamefont {Liao}\ \emph {et~al.}(2021)\citenamefont {Liao},
  \citenamefont {Schraivogel}, \citenamefont {Luo}, \citenamefont {Kats},\ and\
  \citenamefont {Alavi}}]{liao21}%
  \BibitemOpen
  \bibfield  {author} {\bibinfo {author} {\bibfnamefont {K.}~\bibnamefont
  {Liao}}, \bibinfo {author} {\bibfnamefont {T.}~\bibnamefont {Schraivogel}},
  \bibinfo {author} {\bibfnamefont {H.}~\bibnamefont {Luo}}, \bibinfo {author}
  {\bibfnamefont {D.}~\bibnamefont {Kats}}, \ and\ \bibinfo {author}
  {\bibfnamefont {A.}~\bibnamefont {Alavi}},\ }\href {\doibase
  10.1103/PhysRevResearch.3.033072} {\bibfield  {journal} {\bibinfo  {journal}
  {Phys. Rev. Res.}\ }\textbf {\bibinfo {volume} {3}},\ \bibinfo {pages}
  {033072} (\bibinfo {year} {2021})}\BibitemShut {NoStop}%
\bibitem [{\citenamefont {Schraivogel}\ \emph {et~al.}(2021)\citenamefont
  {Schraivogel}, \citenamefont {Cohen}, \citenamefont {Alavi},\ and\
  \citenamefont {Kats}}]{schraivogel21_tc}%
  \BibitemOpen
  \bibfield  {author} {\bibinfo {author} {\bibfnamefont {T.}~\bibnamefont
  {Schraivogel}}, \bibinfo {author} {\bibfnamefont {A.~J.}\ \bibnamefont
  {Cohen}}, \bibinfo {author} {\bibfnamefont {A.}~\bibnamefont {Alavi}}, \ and\
  \bibinfo {author} {\bibfnamefont {D.}~\bibnamefont {Kats}},\ }\href@noop {}
  {\bibfield  {journal} {\bibinfo  {journal} {J. Chem. Phys.}\ }\textbf
  {\bibinfo {volume} {155}},\ \bibinfo {pages} {191101} (\bibinfo {year}
  {2021})}\BibitemShut {NoStop}%
\bibitem [{\citenamefont {Luo}\ and\ \citenamefont {Alavi}(2022)}]{luo22}%
  \BibitemOpen
  \bibfield  {author} {\bibinfo {author} {\bibfnamefont {H.}~\bibnamefont
  {Luo}}\ and\ \bibinfo {author} {\bibfnamefont {A.}~\bibnamefont {Alavi}},\
  }\href {\doibase 10.1063/5.0101776} {\bibfield  {journal} {\bibinfo
  {journal} {J. Chem. Phys.}\ }\textbf {\bibinfo {volume} {157}},\ \bibinfo
  {pages} {074105} (\bibinfo {year} {2022})}\BibitemShut {NoStop}%
\bibitem [{\citenamefont {Motta}\ \emph {et~al.}(2020)\citenamefont {Motta},
  \citenamefont {Gujarati}, \citenamefont {Rice}, \citenamefont {Kumar},
  \citenamefont {Masteran}, \citenamefont {Latone}, \citenamefont {Lee},
  \citenamefont {Valeev},\ and\ \citenamefont {Takeshita}}]{motta20}%
  \BibitemOpen
  \bibfield  {author} {\bibinfo {author} {\bibfnamefont {M.}~\bibnamefont
  {Motta}}, \bibinfo {author} {\bibfnamefont {T.~P.}\ \bibnamefont {Gujarati}},
  \bibinfo {author} {\bibfnamefont {J.~E.}\ \bibnamefont {Rice}}, \bibinfo
  {author} {\bibfnamefont {A.}~\bibnamefont {Kumar}}, \bibinfo {author}
  {\bibfnamefont {C.}~\bibnamefont {Masteran}}, \bibinfo {author}
  {\bibfnamefont {J.~A.}\ \bibnamefont {Latone}}, \bibinfo {author}
  {\bibfnamefont {E.}~\bibnamefont {Lee}}, \bibinfo {author} {\bibfnamefont
  {E.~F.}\ \bibnamefont {Valeev}}, \ and\ \bibinfo {author} {\bibfnamefont
  {T.~Y.}\ \bibnamefont {Takeshita}},\ }\href {\doibase 10.1039/D0CP04106H}
  {\bibfield  {journal} {\bibinfo  {journal} {Phys. Chem. Chem. Phys.}\
  }\textbf {\bibinfo {volume} {22}},\ \bibinfo {pages} {24270} (\bibinfo {year}
  {2020})}\BibitemShut {NoStop}%
\bibitem [{\citenamefont {Baiardi}\ and\ \citenamefont
  {Reiher}(2020)}]{baiardi20}%
  \BibitemOpen
  \bibfield  {author} {\bibinfo {author} {\bibfnamefont {A.}~\bibnamefont
  {Baiardi}}\ and\ \bibinfo {author} {\bibfnamefont {M.}~\bibnamefont
  {Reiher}},\ }\href {\doibase 10.1063/5.0028608} {\bibfield  {journal}
  {\bibinfo  {journal} {J. Chem. Phys.}\ }\textbf {\bibinfo {volume} {153}},\
  \bibinfo {pages} {164115} (\bibinfo {year} {2020})}\BibitemShut {NoStop}%
\bibitem [{\citenamefont {Baiardi}, \citenamefont {Lesiuk},\ and\ \citenamefont
  {Reiher}(2022)}]{baiardi22}%
  \BibitemOpen
  \bibfield  {author} {\bibinfo {author} {\bibfnamefont {A.}~\bibnamefont
  {Baiardi}}, \bibinfo {author} {\bibfnamefont {M.}~\bibnamefont {Lesiuk}}, \
  and\ \bibinfo {author} {\bibfnamefont {M.}~\bibnamefont {Reiher}},\ }\href
  {\doibase 10.1021/acs.jctc.2c00167} {\bibfield  {journal} {\bibinfo
  {journal} {J. Chem. Theory Comput.}\ }\textbf {\bibinfo {volume} {18}},\
  \bibinfo {pages} {4203} (\bibinfo {year} {2022})}\BibitemShut {NoStop}%
\bibitem [{\citenamefont {Khamoshi}\ \emph {et~al.}(2021)\citenamefont
  {Khamoshi}, \citenamefont {Chen}, \citenamefont {Henderson},\ and\
  \citenamefont {Scuseria}}]{khamoshi21}%
  \BibitemOpen
  \bibfield  {author} {\bibinfo {author} {\bibfnamefont {A.}~\bibnamefont
  {Khamoshi}}, \bibinfo {author} {\bibfnamefont {G.~P.}\ \bibnamefont {Chen}},
  \bibinfo {author} {\bibfnamefont {T.~M.}\ \bibnamefont {Henderson}}, \ and\
  \bibinfo {author} {\bibfnamefont {G.~E.}\ \bibnamefont {Scuseria}},\ }\href
  {\doibase 10.1063/5.0039618} {\bibfield  {journal} {\bibinfo  {journal} {J.
  Chem. Phys.}\ }\textbf {\bibinfo {volume} {154}},\ \bibinfo {pages} {074113}
  (\bibinfo {year} {2021})}\BibitemShut {NoStop}%
\bibitem [{\citenamefont {Giner}(2021)}]{giner21}%
  \BibitemOpen
  \bibfield  {author} {\bibinfo {author} {\bibfnamefont {E.}~\bibnamefont
  {Giner}},\ }\href {\doibase 10.1063/5.0044683} {\bibfield  {journal}
  {\bibinfo  {journal} {J. Chem. Phys.}\ }\textbf {\bibinfo {volume} {154}},\
  \bibinfo {pages} {084119} (\bibinfo {year} {2021})}\BibitemShut {NoStop}%
\bibitem [{\citenamefont {Kumar}\ \emph {et~al.}(2022)\citenamefont {Kumar},
  \citenamefont {Asthana}, \citenamefont {Masteran}, \citenamefont {Valeev},
  \citenamefont {Zhang}, \citenamefont {Cincio}, \citenamefont {Tretiak},\ and\
  \citenamefont {Dub}}]{kumar22}%
  \BibitemOpen
  \bibfield  {author} {\bibinfo {author} {\bibfnamefont {A.}~\bibnamefont
  {Kumar}}, \bibinfo {author} {\bibfnamefont {A.}~\bibnamefont {Asthana}},
  \bibinfo {author} {\bibfnamefont {C.}~\bibnamefont {Masteran}}, \bibinfo
  {author} {\bibfnamefont {E.~F.}\ \bibnamefont {Valeev}}, \bibinfo {author}
  {\bibfnamefont {Y.}~\bibnamefont {Zhang}}, \bibinfo {author} {\bibfnamefont
  {L.}~\bibnamefont {Cincio}}, \bibinfo {author} {\bibfnamefont
  {S.}~\bibnamefont {Tretiak}}, \ and\ \bibinfo {author} {\bibfnamefont
  {P.~A.}\ \bibnamefont {Dub}},\ }\href {\doibase 10.1021/acs.jctc.2c00520}
  {\bibfield  {journal} {\bibinfo  {journal} {J. Chem. Theory Comput.}\
  }\textbf {\bibinfo {volume} {18}},\ \bibinfo {pages} {5312} (\bibinfo {year}
  {2022})}\BibitemShut {NoStop}%
\bibitem [{\citenamefont {Dobrautz}\ \emph {et~al.}(2022)\citenamefont
  {Dobrautz}, \citenamefont {Cohen}, \citenamefont {Alavi},\ and\ \citenamefont
  {Giner}}]{dobrautz22}%
  \BibitemOpen
  \bibfield  {author} {\bibinfo {author} {\bibfnamefont {W.}~\bibnamefont
  {Dobrautz}}, \bibinfo {author} {\bibfnamefont {A.~J.}\ \bibnamefont {Cohen}},
  \bibinfo {author} {\bibfnamefont {A.}~\bibnamefont {Alavi}}, \ and\ \bibinfo
  {author} {\bibfnamefont {E.}~\bibnamefont {Giner}},\ }\href@noop {}
  {\bibfield  {journal} {\bibinfo  {journal} {J. Chem. Phys.}\ }\textbf
  {\bibinfo {volume} {156}},\ \bibinfo {pages} {234108} (\bibinfo {year}
  {2022})}\BibitemShut {NoStop}%
\bibitem [{\citenamefont {Ammar}, \citenamefont {Scemama},\ and\ \citenamefont
  {Giner}(2022)}]{ammar22}%
  \BibitemOpen
  \bibfield  {author} {\bibinfo {author} {\bibfnamefont {A.}~\bibnamefont
  {Ammar}}, \bibinfo {author} {\bibfnamefont {A.}~\bibnamefont {Scemama}}, \
  and\ \bibinfo {author} {\bibfnamefont {E.}~\bibnamefont {Giner}},\ }\href
  {\doibase 10.1063/5.0115524} {\bibfield  {journal} {\bibinfo  {journal} {J.
  Chem. Phys.}\ }\textbf {\bibinfo {volume} {157}},\ \bibinfo {pages} {134107}
  (\bibinfo {year} {2022})}\BibitemShut {NoStop}%
\bibitem [{\citenamefont {Liao}\ \emph {et~al.}(2023)\citenamefont {Liao},
  \citenamefont {Zhai}, \citenamefont {Christlmaier}, \citenamefont
  {Schraivogel}, \citenamefont {R\'ios}, \citenamefont {Kats},\ and\
  \citenamefont {Alavi}}]{liao22}%
  \BibitemOpen
  \bibfield  {author} {\bibinfo {author} {\bibfnamefont {K.}~\bibnamefont
  {Liao}}, \bibinfo {author} {\bibfnamefont {H.}~\bibnamefont {Zhai}}, \bibinfo
  {author} {\bibfnamefont {E.~M.}\ \bibnamefont {Christlmaier}}, \bibinfo
  {author} {\bibfnamefont {T.}~\bibnamefont {Schraivogel}}, \bibinfo {author}
  {\bibfnamefont {P.~L.}\ \bibnamefont {R\'ios}}, \bibinfo {author}
  {\bibfnamefont {D.}~\bibnamefont {Kats}}, \ and\ \bibinfo {author}
  {\bibfnamefont {A.}~\bibnamefont {Alavi}},\ }\href
  {https://doi.org/10.1021/acs.jctc.2c01207} {\bibfield  {journal} {\bibinfo
  {journal} {J. Chem. Theory Comput.}\ }\textbf {\bibinfo {volume} {19}},\
  \bibinfo {pages} {1734} (\bibinfo {year} {2023})}\BibitemShut {NoStop}%
\bibitem [{\citenamefont {Ochi}(2023)}]{ochiTC23}%
  \BibitemOpen
  \bibfield  {author} {\bibinfo {author} {\bibfnamefont {M.}~\bibnamefont
  {Ochi}},\ }\href {\doibase 10.1016/j.cpc.2023.108687} {\bibfield  {journal}
  {\bibinfo  {journal} {Comput. Phys. Commun.}\ }\textbf {\bibinfo {volume}
  {287}},\ \bibinfo {pages} {108687} (\bibinfo {year} {2023})}\BibitemShut
  {NoStop}%
\bibitem [{\citenamefont {Lee}\ and\ \citenamefont
  {Thom}(2023)}]{leeStudies2023}%
  \BibitemOpen
  \bibfield  {author} {\bibinfo {author} {\bibfnamefont {N.}~\bibnamefont
  {Lee}}\ and\ \bibinfo {author} {\bibfnamefont {A.~J.~W.}\ \bibnamefont
  {Thom}},\ }\href {\doibase 10.48550/arXiv.2301.02590} {} (\bibinfo {year}
  {2023}),\ \Eprint {http://arxiv.org/abs/arXiv:2301.02590} {arXiv:2301.02590}
  \BibitemShut {NoStop}%
\bibitem [{\citenamefont {Ammar}, \citenamefont {Scemama},\ and\ \citenamefont
  {Giner}(2023)}]{ammarBiorthonormal2023}%
  \BibitemOpen
  \bibfield  {author} {\bibinfo {author} {\bibfnamefont {A.}~\bibnamefont
  {Ammar}}, \bibinfo {author} {\bibfnamefont {A.}~\bibnamefont {Scemama}}, \
  and\ \bibinfo {author} {\bibfnamefont {E.}~\bibnamefont {Giner}},\ }\href
  {\doibase 10.48550/arXiv.2303.02436} {} (\bibinfo {year} {2023}),\ \Eprint
  {http://arxiv.org/abs/arXiv:2303.02436} {arXiv:2303.02436} \BibitemShut
  {NoStop}%
\bibitem [{\citenamefont {Haupt}\ \emph {et~al.}(2023)\citenamefont {Haupt},
  \citenamefont {Hosseini}, \citenamefont {R\'ios}, \citenamefont {Dobrautz},
  \citenamefont {Cohen},\ and\ \citenamefont {Alavi}}]{haupt23}%
  \BibitemOpen
  \bibfield  {author} {\bibinfo {author} {\bibfnamefont {J.~P.}\ \bibnamefont
  {Haupt}}, \bibinfo {author} {\bibfnamefont {S.~M.}\ \bibnamefont {Hosseini}},
  \bibinfo {author} {\bibfnamefont {P.~L.}\ \bibnamefont {R\'ios}}, \bibinfo
  {author} {\bibfnamefont {W.}~\bibnamefont {Dobrautz}}, \bibinfo {author}
  {\bibfnamefont {A.}~\bibnamefont {Cohen}}, \ and\ \bibinfo {author}
  {\bibfnamefont {A.}~\bibnamefont {Alavi}},\ }\href
  {https://doi.org/10.48550/arXiv.2302.13683} {} (\bibinfo {year} {2023}),\
  \Eprint {http://arxiv.org/abs/arXiv:2302.13683} {arXiv:2302.13683}
  \BibitemShut {NoStop}%
\bibitem [{\citenamefont {Shiozaki}\ and\ \citenamefont
  {Werner}(2013)}]{shiozaki_multireference_2013}%
  \BibitemOpen
  \bibfield  {author} {\bibinfo {author} {\bibfnamefont {T.}~\bibnamefont
  {Shiozaki}}\ and\ \bibinfo {author} {\bibfnamefont {H.-J.}\ \bibnamefont
  {Werner}},\ }\href {\doibase 10.1080/00268976.2013.779393} {\bibfield
  {journal} {\bibinfo  {journal} {Mol. Phys.}\ }\textbf {\bibinfo {volume}
  {111}},\ \bibinfo {pages} {607} (\bibinfo {year} {2013})}\BibitemShut
  {NoStop}%
\bibitem [{\citenamefont {Paldus}(2017)}]{paldus17}%
  \BibitemOpen
  \bibfield  {author} {\bibinfo {author} {\bibfnamefont {J.}~\bibnamefont
  {Paldus}},\ }\href {\doibase 10.1007/s10910-016-0688-6} {\bibfield  {journal}
  {\bibinfo  {journal} {J. Mat. Chem.}\ }\textbf {\bibinfo {volume} {55}},\
  \bibinfo {pages} {477} (\bibinfo {year} {2017})}\BibitemShut {NoStop}%
\bibitem [{\citenamefont {Jankowski}\ and\ \citenamefont
  {Paldus}(1980)}]{jankowski80}%
  \BibitemOpen
  \bibfield  {author} {\bibinfo {author} {\bibfnamefont {K.}~\bibnamefont
  {Jankowski}}\ and\ \bibinfo {author} {\bibfnamefont {J.}~\bibnamefont
  {Paldus}},\ }\href {\doibase 10.1002/qua.560180511} {\bibfield  {journal}
  {\bibinfo  {journal} {Int. J. Quantum Chem.}\ }\textbf {\bibinfo {volume}
  {18}},\ \bibinfo {pages} {1243} (\bibinfo {year} {1980})}\BibitemShut
  {NoStop}%
\bibitem [{\citenamefont {Chiles}\ and\ \citenamefont
  {Dykstra}(1981)}]{chiles81}%
  \BibitemOpen
  \bibfield  {author} {\bibinfo {author} {\bibfnamefont {R.~A.}\ \bibnamefont
  {Chiles}}\ and\ \bibinfo {author} {\bibfnamefont {C.~E.}\ \bibnamefont
  {Dykstra}},\ }\href@noop {} {\bibfield  {journal} {\bibinfo  {journal} {Chem.
  Phys. Lett.}\ }\textbf {\bibinfo {volume} {80}},\ \bibinfo {pages} {4}
  (\bibinfo {year} {1981})}\BibitemShut {NoStop}%
\bibitem [{\citenamefont {Paldus}, \citenamefont {\v{C}\'{\i}\v{z}ek},\ and\
  \citenamefont {Takahashi}(1984)}]{paldus84}%
  \BibitemOpen
  \bibfield  {author} {\bibinfo {author} {\bibfnamefont {J.}~\bibnamefont
  {Paldus}}, \bibinfo {author} {\bibfnamefont {J.}~\bibnamefont
  {\v{C}\'{\i}\v{z}ek}}, \ and\ \bibinfo {author} {\bibfnamefont
  {M.}~\bibnamefont {Takahashi}},\ }\href {\doibase 10.1103/PhysRevA.30.2193}
  {\bibfield  {journal} {\bibinfo  {journal} {Phys. Rev. A}\ }\textbf {\bibinfo
  {volume} {30}},\ \bibinfo {pages} {2193} (\bibinfo {year}
  {1984})}\BibitemShut {NoStop}%
\bibitem [{\citenamefont {Bartlett}\ and\ \citenamefont
  {Musia\l}(2006)}]{bartlett06}%
  \BibitemOpen
  \bibfield  {author} {\bibinfo {author} {\bibfnamefont {R.~J.}\ \bibnamefont
  {Bartlett}}\ and\ \bibinfo {author} {\bibfnamefont {M.}~\bibnamefont
  {Musia\l}},\ }\href {\doibase 10.1063/1.2387952} {\bibfield  {journal}
  {\bibinfo  {journal} {J. Chem. Phys.}\ }\textbf {\bibinfo {volume} {125}},\
  \bibinfo {pages} {204105} (\bibinfo {year} {2006})}\BibitemShut {NoStop}%
\bibitem [{\citenamefont {Musia\l}\ and\ \citenamefont
  {Bartlett}(2007)}]{musial07}%
  \BibitemOpen
  \bibfield  {author} {\bibinfo {author} {\bibfnamefont {M.}~\bibnamefont
  {Musia\l}}\ and\ \bibinfo {author} {\bibfnamefont {R.~J.}\ \bibnamefont
  {Bartlett}},\ }\href {\doibase 10.1063/1.2747245} {\bibfield  {journal}
  {\bibinfo  {journal} {J. Chem. Phys.}\ }\textbf {\bibinfo {volume} {127}},\
  \bibinfo {pages} {024106} (\bibinfo {year} {2007})}\BibitemShut {NoStop}%
\bibitem [{\citenamefont {Huntington}\ and\ \citenamefont
  {Nooijen}(2010)}]{huntington10}%
  \BibitemOpen
  \bibfield  {author} {\bibinfo {author} {\bibfnamefont {L.~M.~J.}\
  \bibnamefont {Huntington}}\ and\ \bibinfo {author} {\bibfnamefont
  {M.}~\bibnamefont {Nooijen}},\ }\href {\doibase doi:10.1063/1.3494113}
  {\bibfield  {journal} {\bibinfo  {journal} {J. Chem. Phys.}\ }\textbf
  {\bibinfo {volume} {133}},\ \bibinfo {pages} {184109} (\bibinfo {year}
  {2010})}\BibitemShut {NoStop}%
\bibitem [{\citenamefont {Huntington}\ \emph {et~al.}(2012)\citenamefont
  {Huntington}, \citenamefont {Hansen}, \citenamefont {Neese},\ and\
  \citenamefont {Nooijen}}]{huntington12}%
  \BibitemOpen
  \bibfield  {author} {\bibinfo {author} {\bibfnamefont {L.~M.~J.}\
  \bibnamefont {Huntington}}, \bibinfo {author} {\bibfnamefont
  {A.}~\bibnamefont {Hansen}}, \bibinfo {author} {\bibfnamefont
  {F.}~\bibnamefont {Neese}}, \ and\ \bibinfo {author} {\bibfnamefont
  {M.}~\bibnamefont {Nooijen}},\ }\href {\doibase 10.1063/1.3682325} {\bibfield
   {journal} {\bibinfo  {journal} {J. Chem. Phys.}\ }\textbf {\bibinfo {volume}
  {136}},\ \bibinfo {pages} {064101} (\bibinfo {year} {2012})}\BibitemShut
  {NoStop}%
\bibitem [{\citenamefont {Kats}\ and\ \citenamefont {Manby}(2013)}]{kats13}%
  \BibitemOpen
  \bibfield  {author} {\bibinfo {author} {\bibfnamefont {D.}~\bibnamefont
  {Kats}}\ and\ \bibinfo {author} {\bibfnamefont {F.~R.}\ \bibnamefont
  {Manby}},\ }\href {\doibase 10.1063/1.4813481} {\bibfield  {journal}
  {\bibinfo  {journal} {J. Chem. Phys.}\ }\textbf {\bibinfo {volume} {139}},\
  \bibinfo {pages} {021102} (\bibinfo {year} {2013})}\BibitemShut {NoStop}%
\bibitem [{\citenamefont {Kats}(2014)}]{kats14}%
  \BibitemOpen
  \bibfield  {author} {\bibinfo {author} {\bibfnamefont {D.}~\bibnamefont
  {Kats}},\ }\href {\doibase 10.1063/1.4892792} {\bibfield  {journal} {\bibinfo
   {journal} {J. Chem. Phys.}\ }\textbf {\bibinfo {volume} {141}},\ \bibinfo
  {pages} {061101} (\bibinfo {year} {2014})}\BibitemShut {NoStop}%
\bibitem [{\citenamefont {Kats}\ \emph {et~al.}(2015)\citenamefont {Kats},
  \citenamefont {Kreplin}, \citenamefont {Werner},\ and\ \citenamefont
  {Manby}}]{kats15}%
  \BibitemOpen
  \bibfield  {author} {\bibinfo {author} {\bibfnamefont {D.}~\bibnamefont
  {Kats}}, \bibinfo {author} {\bibfnamefont {D.}~\bibnamefont {Kreplin}},
  \bibinfo {author} {\bibfnamefont {H.-J.}\ \bibnamefont {Werner}}, \ and\
  \bibinfo {author} {\bibfnamefont {F.~R.}\ \bibnamefont {Manby}},\ }\href
  {\doibase 10.1063/1.4907591} {\bibfield  {journal} {\bibinfo  {journal} {J.
  Chem. Phys.}\ }\textbf {\bibinfo {volume} {142}},\ \bibinfo {pages} {064111}
  (\bibinfo {year} {2015})}\BibitemShut {NoStop}%
\bibitem [{\citenamefont {Kats}(2016)}]{kats16}%
  \BibitemOpen
  \bibfield  {author} {\bibinfo {author} {\bibfnamefont {D.}~\bibnamefont
  {Kats}},\ }\href {\doibase 10.1063/1.4940398} {\bibfield  {journal} {\bibinfo
   {journal} {J. Chem. Phys.}\ }\textbf {\bibinfo {volume} {144}},\ \bibinfo
  {pages} {044102} (\bibinfo {year} {2016})}\BibitemShut {NoStop}%
\bibitem [{\citenamefont {Rishi}\ \emph {et~al.}(2017)\citenamefont {Rishi},
  \citenamefont {Perera}, \citenamefont {Nooijen},\ and\ \citenamefont
  {Bartlett}}]{rishi17}%
  \BibitemOpen
  \bibfield  {author} {\bibinfo {author} {\bibfnamefont {V.}~\bibnamefont
  {Rishi}}, \bibinfo {author} {\bibfnamefont {A.}~\bibnamefont {Perera}},
  \bibinfo {author} {\bibfnamefont {M.}~\bibnamefont {Nooijen}}, \ and\
  \bibinfo {author} {\bibfnamefont {R.~J.}\ \bibnamefont {Bartlett}},\ }\href
  {\doibase 10.1063/1.4979078} {\bibfield  {journal} {\bibinfo  {journal} {J.
  Chem. Phys.}\ }\textbf {\bibinfo {volume} {146}},\ \bibinfo {pages} {144104}
  (\bibinfo {year} {2017})}\BibitemShut {NoStop}%
\bibitem [{\citenamefont {Kats}, \citenamefont {Usvyat},\ and\ \citenamefont
  {Manby}(2018)}]{kats18}%
  \BibitemOpen
  \bibfield  {author} {\bibinfo {author} {\bibfnamefont {D.}~\bibnamefont
  {Kats}}, \bibinfo {author} {\bibfnamefont {D.}~\bibnamefont {Usvyat}}, \ and\
  \bibinfo {author} {\bibfnamefont {F.~R.}\ \bibnamefont {Manby}},\ }\href
  {\doibase 10.1080/00268976.2018.1448947} {\bibfield  {journal} {\bibinfo
  {journal} {Mol. Phys.}\ }\textbf {\bibinfo {volume} {116}},\ \bibinfo {pages}
  {1496} (\bibinfo {year} {2018})}\BibitemShut {NoStop}%
\bibitem [{\citenamefont {Kats}\ and\ \citenamefont
  {K{\"o}hn}(2019)}]{kats19_dc}%
  \BibitemOpen
  \bibfield  {author} {\bibinfo {author} {\bibfnamefont {D.}~\bibnamefont
  {Kats}}\ and\ \bibinfo {author} {\bibfnamefont {A.}~\bibnamefont
  {K{\"o}hn}},\ }\href {\doibase 10.1063/1.5096343} {\bibfield  {journal}
  {\bibinfo  {journal} {J. Chem. Phys.}\ }\textbf {\bibinfo {volume} {150}},\
  \bibinfo {pages} {151101} (\bibinfo {year} {2019})}\BibitemShut {NoStop}%
\bibitem [{\citenamefont {Rishi}\ and\ \citenamefont {Valeev}(2019)}]{rishi19}%
  \BibitemOpen
  \bibfield  {author} {\bibinfo {author} {\bibfnamefont {V.}~\bibnamefont
  {Rishi}}\ and\ \bibinfo {author} {\bibfnamefont {E.~F.}\ \bibnamefont
  {Valeev}},\ }\href {\doibase 10.1063/1.5097150} {\bibfield  {journal}
  {\bibinfo  {journal} {J. Chem. Phys.}\ }\textbf {\bibinfo {volume} {151}},\
  \bibinfo {pages} {064102} (\bibinfo {year} {2019})}\BibitemShut {NoStop}%
\bibitem [{\citenamefont {Schraivogel}\ and\ \citenamefont
  {Kats}(2021)}]{schraivogel21_dc}%
  \BibitemOpen
  \bibfield  {author} {\bibinfo {author} {\bibfnamefont {T.}~\bibnamefont
  {Schraivogel}}\ and\ \bibinfo {author} {\bibfnamefont {D.}~\bibnamefont
  {Kats}},\ }\href {\doibase 10.1063/5.0059181} {\bibfield  {journal} {\bibinfo
   {journal} {J. Chem. Phys.}\ }\textbf {\bibinfo {volume} {155}},\ \bibinfo
  {pages} {064101} (\bibinfo {year} {2021})}\BibitemShut {NoStop}%
\bibitem [{\citenamefont {Schmidt}\ and\ \citenamefont
  {Moskowitz}(1990)}]{schmidt90}%
  \BibitemOpen
  \bibfield  {author} {\bibinfo {author} {\bibfnamefont {K.~E.}\ \bibnamefont
  {Schmidt}}\ and\ \bibinfo {author} {\bibfnamefont {J.~W.}\ \bibnamefont
  {Moskowitz}},\ }\href {\doibase 10.1063/1.458750} {\bibfield  {journal}
  {\bibinfo  {journal} {J. Chem. Phys.}\ }\textbf {\bibinfo {volume} {93}},\
  \bibinfo {pages} {4172} (\bibinfo {year} {1990})}\BibitemShut {NoStop}%
\bibitem [{\citenamefont {Koch}\ \emph {et~al.}(1994)\citenamefont {Koch},
  \citenamefont {Christiansen}, \citenamefont {Kobayashi}, \citenamefont
  {J{\o}rgensen},\ and\ \citenamefont {Helgaker}}]{koch94}%
  \BibitemOpen
  \bibfield  {author} {\bibinfo {author} {\bibfnamefont {H.}~\bibnamefont
  {Koch}}, \bibinfo {author} {\bibfnamefont {O.}~\bibnamefont {Christiansen}},
  \bibinfo {author} {\bibfnamefont {R.}~\bibnamefont {Kobayashi}}, \bibinfo
  {author} {\bibfnamefont {P.}~\bibnamefont {J{\o}rgensen}}, \ and\ \bibinfo
  {author} {\bibfnamefont {T.}~\bibnamefont {Helgaker}},\ }\href {\doibase
  10.1016/0009-2614(94)00898-1} {\bibfield  {journal} {\bibinfo  {journal}
  {Chem. Phys. Lett.}\ }\textbf {\bibinfo {volume} {228}},\ \bibinfo {pages}
  {233} (\bibinfo {year} {1994})}\BibitemShut {NoStop}%
\bibitem [{\citenamefont {Kats}\ and\ \citenamefont
  {Schraivogel}(2022)}]{quantwo}%
  \BibitemOpen
  \bibfield  {author} {\bibinfo {author} {\bibfnamefont {D.}~\bibnamefont
  {Kats}}\ and\ \bibinfo {author} {\bibfnamefont {T.}~\bibnamefont
  {Schraivogel}},\ }\href@noop {} {\enquote {\bibinfo {title} {Quantwo:
  second-quantization program},}\ } (\bibinfo {year} {2022})\BibitemShut
  {NoStop}%
\bibitem [{\citenamefont {Shamasundar}, \citenamefont {Knizia},\ and\
  \citenamefont {Werner}(2011)}]{shamasundar2011}%
  \BibitemOpen
  \bibfield  {author} {\bibinfo {author} {\bibfnamefont {K.~R.}\ \bibnamefont
  {Shamasundar}}, \bibinfo {author} {\bibfnamefont {G.}~\bibnamefont {Knizia}},
  \ and\ \bibinfo {author} {\bibfnamefont {H.-J.}\ \bibnamefont {Werner}},\
  }\href {\doibase 10.1063/1.3609809} {\bibfield  {journal} {\bibinfo
  {journal} {J. Chem. Phys.}\ }\textbf {\bibinfo {volume} {135}},\ \bibinfo
  {pages} {054101} (\bibinfo {year} {2011})}\BibitemShut {NoStop}%
\bibitem [{\citenamefont {Werner}\ \emph {et~al.}(2012)\citenamefont {Werner},
  \citenamefont {Knowles}, \citenamefont {Knizia}, \citenamefont {Manby},\ and\
  \citenamefont {Sch{\"u}tz}}]{MOLPRO-WIREs}%
  \BibitemOpen
  \bibfield  {author} {\bibinfo {author} {\bibfnamefont {H.-J.}\ \bibnamefont
  {Werner}}, \bibinfo {author} {\bibfnamefont {P.~J.}\ \bibnamefont {Knowles}},
  \bibinfo {author} {\bibfnamefont {G.}~\bibnamefont {Knizia}}, \bibinfo
  {author} {\bibfnamefont {F.~R.}\ \bibnamefont {Manby}}, \ and\ \bibinfo
  {author} {\bibfnamefont {M.}~\bibnamefont {Sch{\"u}tz}},\ }\href@noop {}
  {\bibfield  {journal} {\bibinfo  {journal} {WIREs Comput. Mol. Sci.}\
  }\textbf {\bibinfo {volume} {2}},\ \bibinfo {pages} {242} (\bibinfo {year}
  {2012})}\BibitemShut {NoStop}%
\bibitem [{\citenamefont {Werner}\ \emph {et~al.}(2020)\citenamefont {Werner},
  \citenamefont {Knowles}, \citenamefont {Manby}, \citenamefont {Black},
  \citenamefont {Doll}, \citenamefont {Heßelmann}, \citenamefont {Kats},
  \citenamefont {Köhn}, \citenamefont {Korona}, \citenamefont {Kreplin},
  \citenamefont {Ma}, \citenamefont {Miller}, \citenamefont {Mitrushchenkov},
  \citenamefont {Peterson}, \citenamefont {Polyak}, \citenamefont {Rauhut},\
  and\ \citenamefont {Sibaev}}]{MOLPRO-JCP}%
  \BibitemOpen
  \bibfield  {author} {\bibinfo {author} {\bibfnamefont {H.-J.}\ \bibnamefont
  {Werner}}, \bibinfo {author} {\bibfnamefont {P.~J.}\ \bibnamefont {Knowles}},
  \bibinfo {author} {\bibfnamefont {F.~R.}\ \bibnamefont {Manby}}, \bibinfo
  {author} {\bibfnamefont {J.~A.}\ \bibnamefont {Black}}, \bibinfo {author}
  {\bibfnamefont {K.}~\bibnamefont {Doll}}, \bibinfo {author} {\bibfnamefont
  {A.}~\bibnamefont {Heßelmann}}, \bibinfo {author} {\bibfnamefont
  {D.}~\bibnamefont {Kats}}, \bibinfo {author} {\bibfnamefont {A.}~\bibnamefont
  {Köhn}}, \bibinfo {author} {\bibfnamefont {T.}~\bibnamefont {Korona}},
  \bibinfo {author} {\bibfnamefont {D.~A.}\ \bibnamefont {Kreplin}}, \bibinfo
  {author} {\bibfnamefont {Q.}~\bibnamefont {Ma}}, \bibinfo {author}
  {\bibfnamefont {T.~F.}\ \bibnamefont {Miller}}, \bibinfo {author}
  {\bibfnamefont {A.}~\bibnamefont {Mitrushchenkov}}, \bibinfo {author}
  {\bibfnamefont {K.~A.}\ \bibnamefont {Peterson}}, \bibinfo {author}
  {\bibfnamefont {I.}~\bibnamefont {Polyak}}, \bibinfo {author} {\bibfnamefont
  {G.}~\bibnamefont {Rauhut}}, \ and\ \bibinfo {author} {\bibfnamefont
  {M.}~\bibnamefont {Sibaev}},\ }\href {\doibase 10.1063/5.0005081} {\bibfield
  {journal} {\bibinfo  {journal} {J. Chem. Phys.}\ }\textbf {\bibinfo {volume}
  {152}},\ \bibinfo {pages} {144107} (\bibinfo {year} {2020})}\BibitemShut
  {NoStop}%
\bibitem [{\citenamefont {Werner}\ \emph {et~al.}()\citenamefont {Werner},
  \citenamefont {Knowles}, \citenamefont {Knizia}, \citenamefont {Manby},
  \citenamefont {{Sch\"{u}tz}}, \citenamefont {Celani}, \citenamefont
  {Gy\"orffy}, \citenamefont {Kats}, \citenamefont {Korona}, \citenamefont
  {Lindh}, \citenamefont {Mitrushenkov}, \citenamefont {Rauhut}, \citenamefont
  {Shamasundar}, \citenamefont {Adler}, \citenamefont {Amos}, \citenamefont
  {Bennie}, \citenamefont {Bernhardsson}, \citenamefont {Berning},
  \citenamefont {Cooper}, \citenamefont {Deegan}, \citenamefont {Dobbyn},
  \citenamefont {Eckert}, \citenamefont {Goll}, \citenamefont {Hampel},
  \citenamefont {Hesselmann}, \citenamefont {Hetzer}, \citenamefont {Hrenar},
  \citenamefont {Jansen}, \citenamefont {K\"oppl}, \citenamefont {Lee},
  \citenamefont {Liu}, \citenamefont {Lloyd}, \citenamefont {Ma}, \citenamefont
  {Mata}, \citenamefont {May}, \citenamefont {McNicholas}, \citenamefont
  {Meyer}, \citenamefont {{Miller III}}, \citenamefont {Mura}, \citenamefont
  {Nicklass}, \citenamefont {O'Neill}, \citenamefont {Palmieri}, \citenamefont
  {Peng}, \citenamefont {Pfl\"uger}, \citenamefont {Pitzer}, \citenamefont
  {Reiher}, \citenamefont {Shiozaki}, \citenamefont {Stoll}, \citenamefont
  {Stone}, \citenamefont {Tarroni}, \citenamefont {Thorsteinsson},
  \citenamefont {Wang},\ and\ \citenamefont {Welborn}}]{MOLPRO}%
  \BibitemOpen
  \bibfield  {author} {\bibinfo {author} {\bibfnamefont {H.-J.}\ \bibnamefont
  {Werner}}, \bibinfo {author} {\bibfnamefont {P.~J.}\ \bibnamefont {Knowles}},
  \bibinfo {author} {\bibfnamefont {G.}~\bibnamefont {Knizia}}, \bibinfo
  {author} {\bibfnamefont {F.~R.}\ \bibnamefont {Manby}}, \bibinfo {author}
  {\bibfnamefont {M.}~\bibnamefont {{Sch\"{u}tz}}}, \bibinfo {author}
  {\bibfnamefont {P.}~\bibnamefont {Celani}}, \bibinfo {author} {\bibfnamefont
  {W.}~\bibnamefont {Gy\"orffy}}, \bibinfo {author} {\bibfnamefont
  {D.}~\bibnamefont {Kats}}, \bibinfo {author} {\bibfnamefont {T.}~\bibnamefont
  {Korona}}, \bibinfo {author} {\bibfnamefont {R.}~\bibnamefont {Lindh}},
  \bibinfo {author} {\bibfnamefont {A.}~\bibnamefont {Mitrushenkov}}, \bibinfo
  {author} {\bibfnamefont {G.}~\bibnamefont {Rauhut}}, \bibinfo {author}
  {\bibfnamefont {K.~R.}\ \bibnamefont {Shamasundar}}, \bibinfo {author}
  {\bibfnamefont {T.~B.}\ \bibnamefont {Adler}}, \bibinfo {author}
  {\bibfnamefont {R.~D.}\ \bibnamefont {Amos}}, \bibinfo {author}
  {\bibfnamefont {S.~J.}\ \bibnamefont {Bennie}}, \bibinfo {author}
  {\bibfnamefont {A.}~\bibnamefont {Bernhardsson}}, \bibinfo {author}
  {\bibfnamefont {A.}~\bibnamefont {Berning}}, \bibinfo {author} {\bibfnamefont
  {D.~L.}\ \bibnamefont {Cooper}}, \bibinfo {author} {\bibfnamefont {M.~J.~O.}\
  \bibnamefont {Deegan}}, \bibinfo {author} {\bibfnamefont {A.~J.}\
  \bibnamefont {Dobbyn}}, \bibinfo {author} {\bibfnamefont {F.}~\bibnamefont
  {Eckert}}, \bibinfo {author} {\bibfnamefont {E.}~\bibnamefont {Goll}},
  \bibinfo {author} {\bibfnamefont {C.}~\bibnamefont {Hampel}}, \bibinfo
  {author} {\bibfnamefont {A.}~\bibnamefont {Hesselmann}}, \bibinfo {author}
  {\bibfnamefont {G.}~\bibnamefont {Hetzer}}, \bibinfo {author} {\bibfnamefont
  {T.}~\bibnamefont {Hrenar}}, \bibinfo {author} {\bibfnamefont
  {G.}~\bibnamefont {Jansen}}, \bibinfo {author} {\bibfnamefont
  {C.}~\bibnamefont {K\"oppl}}, \bibinfo {author} {\bibfnamefont {S.~J.~R.}\
  \bibnamefont {Lee}}, \bibinfo {author} {\bibfnamefont {Y.}~\bibnamefont
  {Liu}}, \bibinfo {author} {\bibfnamefont {A.~W.}\ \bibnamefont {Lloyd}},
  \bibinfo {author} {\bibfnamefont {Q.}~\bibnamefont {Ma}}, \bibinfo {author}
  {\bibfnamefont {R.~A.}\ \bibnamefont {Mata}}, \bibinfo {author}
  {\bibfnamefont {A.~J.}\ \bibnamefont {May}}, \bibinfo {author} {\bibfnamefont
  {S.~J.}\ \bibnamefont {McNicholas}}, \bibinfo {author} {\bibfnamefont
  {W.}~\bibnamefont {Meyer}}, \bibinfo {author} {\bibfnamefont {T.~F.}\
  \bibnamefont {{Miller III}}}, \bibinfo {author} {\bibfnamefont {M.~E.}\
  \bibnamefont {Mura}}, \bibinfo {author} {\bibfnamefont {A.}~\bibnamefont
  {Nicklass}}, \bibinfo {author} {\bibfnamefont {D.~P.}\ \bibnamefont
  {O'Neill}}, \bibinfo {author} {\bibfnamefont {P.}~\bibnamefont {Palmieri}},
  \bibinfo {author} {\bibfnamefont {D.}~\bibnamefont {Peng}}, \bibinfo {author}
  {\bibfnamefont {K.}~\bibnamefont {Pfl\"uger}}, \bibinfo {author}
  {\bibfnamefont {R.}~\bibnamefont {Pitzer}}, \bibinfo {author} {\bibfnamefont
  {M.}~\bibnamefont {Reiher}}, \bibinfo {author} {\bibfnamefont
  {T.}~\bibnamefont {Shiozaki}}, \bibinfo {author} {\bibfnamefont
  {H.}~\bibnamefont {Stoll}}, \bibinfo {author} {\bibfnamefont {A.~J.}\
  \bibnamefont {Stone}}, \bibinfo {author} {\bibfnamefont {R.}~\bibnamefont
  {Tarroni}}, \bibinfo {author} {\bibfnamefont {T.}~\bibnamefont
  {Thorsteinsson}}, \bibinfo {author} {\bibfnamefont {M.}~\bibnamefont {Wang}},
  \ and\ \bibinfo {author} {\bibfnamefont {M.}~\bibnamefont {Welborn}},\
  }\href@noop {} {\enquote {\bibinfo {title} {Molpro, 2022.2, a package of ab
  initio programs},}\ }\bibinfo {note} {See https://www.molpro.net}\BibitemShut
  {NoStop}%
\bibitem [{\citenamefont {Dunning}(1989)}]{dunning89}%
  \BibitemOpen
  \bibfield  {author} {\bibinfo {author} {\bibfnamefont {T.~H.}\ \bibnamefont
  {Dunning}},\ }\href {\doibase 10.1063/1.456153} {\bibfield  {journal}
  {\bibinfo  {journal} {J. Chem. Phys.}\ }\textbf {\bibinfo {volume} {90}},\
  \bibinfo {pages} {1007} (\bibinfo {year} {1989})}\BibitemShut {NoStop}%
\bibitem [{\citenamefont {Foulkes}\ \emph {et~al.}(2001)\citenamefont
  {Foulkes}, \citenamefont {Mitas}, \citenamefont {Needs},\ and\ \citenamefont
  {Rajagopal}}]{foulkes01}%
  \BibitemOpen
  \bibfield  {author} {\bibinfo {author} {\bibfnamefont {W.~M.~C.}\
  \bibnamefont {Foulkes}}, \bibinfo {author} {\bibfnamefont {L.}~\bibnamefont
  {Mitas}}, \bibinfo {author} {\bibfnamefont {R.~J.}\ \bibnamefont {Needs}}, \
  and\ \bibinfo {author} {\bibfnamefont {G.}~\bibnamefont {Rajagopal}},\ }\href
  {\doibase 10.1103/RevModPhys.73.33} {\bibfield  {journal} {\bibinfo
  {journal} {Rev. Mod. Phys.}\ }\textbf {\bibinfo {volume} {73}},\ \bibinfo
  {pages} {33} (\bibinfo {year} {2001})}\BibitemShut {NoStop}%
\bibitem [{\citenamefont {Drummond}, \citenamefont {Towler},\ and\
  \citenamefont {Needs}(2004)}]{drummond04}%
  \BibitemOpen
  \bibfield  {author} {\bibinfo {author} {\bibfnamefont {N.~D.}\ \bibnamefont
  {Drummond}}, \bibinfo {author} {\bibfnamefont {M.~D.}\ \bibnamefont
  {Towler}}, \ and\ \bibinfo {author} {\bibfnamefont {R.~J.}\ \bibnamefont
  {Needs}},\ }\href {\doibase 10.1103/PhysRevB.70.235119} {\bibfield  {journal}
  {\bibinfo  {journal} {Phys. Rev. B}\ }\textbf {\bibinfo {volume} {70}},\
  \bibinfo {pages} {235119} (\bibinfo {year} {2004})}\BibitemShut {NoStop}%
\bibitem [{\citenamefont {Umrigar}, \citenamefont {Wilson},\ and\ \citenamefont
  {Wilkins}(1988)}]{umrigar88}%
  \BibitemOpen
  \bibfield  {author} {\bibinfo {author} {\bibfnamefont {C.~J.}\ \bibnamefont
  {Umrigar}}, \bibinfo {author} {\bibfnamefont {K.~G.}\ \bibnamefont {Wilson}},
  \ and\ \bibinfo {author} {\bibfnamefont {J.~W.}\ \bibnamefont {Wilkins}},\
  }\href {\doibase 10.1103/PhysRevLett.60.1719} {\bibfield  {journal} {\bibinfo
   {journal} {Phys. Rev. Lett.}\ }\textbf {\bibinfo {volume} {60}},\ \bibinfo
  {pages} {1719} (\bibinfo {year} {1988})}\BibitemShut {NoStop}%
\bibitem [{\citenamefont {Kent}, \citenamefont {Needs},\ and\ \citenamefont
  {Rajagopal}(1999)}]{kent99}%
  \BibitemOpen
  \bibfield  {author} {\bibinfo {author} {\bibfnamefont {P.~R.~C.}\
  \bibnamefont {Kent}}, \bibinfo {author} {\bibfnamefont {R.~J.}\ \bibnamefont
  {Needs}}, \ and\ \bibinfo {author} {\bibfnamefont {G.}~\bibnamefont
  {Rajagopal}},\ }\href {\doibase 10.1103/PhysRevB.59.12344} {\bibfield
  {journal} {\bibinfo  {journal} {Phys. Rev. B}\ }\textbf {\bibinfo {volume}
  {59}},\ \bibinfo {pages} {12344} (\bibinfo {year} {1999})}\BibitemShut
  {NoStop}%
\bibitem [{\citenamefont {Needs}\ \emph {et~al.}(2020)\citenamefont {Needs},
  \citenamefont {Towler}, \citenamefont {Drummond}, \citenamefont
  {L\'opez~R\'ios},\ and\ \citenamefont {Trail}}]{needs20}%
  \BibitemOpen
  \bibfield  {author} {\bibinfo {author} {\bibfnamefont {R.~J.}\ \bibnamefont
  {Needs}}, \bibinfo {author} {\bibfnamefont {M.~D.}\ \bibnamefont {Towler}},
  \bibinfo {author} {\bibfnamefont {N.~D.}\ \bibnamefont {Drummond}}, \bibinfo
  {author} {\bibfnamefont {P.}~\bibnamefont {L\'opez~R\'ios}}, \ and\ \bibinfo
  {author} {\bibfnamefont {J.~R.}\ \bibnamefont {Trail}},\ }\href {\doibase
  10.1063/1.5144288} {\bibfield  {journal} {\bibinfo  {journal} {J. Chem.
  Phys.}\ }\textbf {\bibinfo {volume} {152}},\ \bibinfo {pages} {154106}
  (\bibinfo {year} {2020})}\BibitemShut {NoStop}%
\bibitem [{\citenamefont {Sun}(2015)}]{sun15}%
  \BibitemOpen
  \bibfield  {author} {\bibinfo {author} {\bibfnamefont {Q.}~\bibnamefont
  {Sun}},\ }\href {\doibase 10.1002/jcc.23981} {\bibfield  {journal} {\bibinfo
  {journal} {J. Comput. Chem.}\ }\textbf {\bibinfo {volume} {36}},\ \bibinfo
  {pages} {1664} (\bibinfo {year} {2015})}\BibitemShut {NoStop}%
\bibitem [{\citenamefont {Sun}\ \emph {et~al.}(2018)\citenamefont {Sun},
  \citenamefont {Berkelbach}, \citenamefont {Blunt}, \citenamefont {Booth},
  \citenamefont {Guo}, \citenamefont {Li}, \citenamefont {Liu}, \citenamefont
  {McClain}, \citenamefont {Sayfutyarova}, \citenamefont {Sharma},
  \citenamefont {Wouters},\ and\ \citenamefont {Chan}}]{sun18}%
  \BibitemOpen
  \bibfield  {author} {\bibinfo {author} {\bibfnamefont {Q.}~\bibnamefont
  {Sun}}, \bibinfo {author} {\bibfnamefont {T.~C.}\ \bibnamefont {Berkelbach}},
  \bibinfo {author} {\bibfnamefont {N.~S.}\ \bibnamefont {Blunt}}, \bibinfo
  {author} {\bibfnamefont {G.~H.}\ \bibnamefont {Booth}}, \bibinfo {author}
  {\bibfnamefont {S.}~\bibnamefont {Guo}}, \bibinfo {author} {\bibfnamefont
  {Z.}~\bibnamefont {Li}}, \bibinfo {author} {\bibfnamefont {J.}~\bibnamefont
  {Liu}}, \bibinfo {author} {\bibfnamefont {J.~D.}\ \bibnamefont {McClain}},
  \bibinfo {author} {\bibfnamefont {E.~R.}\ \bibnamefont {Sayfutyarova}},
  \bibinfo {author} {\bibfnamefont {S.}~\bibnamefont {Sharma}}, \bibinfo
  {author} {\bibfnamefont {S.}~\bibnamefont {Wouters}}, \ and\ \bibinfo
  {author} {\bibfnamefont {G.~K.}\ \bibnamefont {Chan}},\ }\href {\doibase
  10.1002/wcms.1340} {\bibfield  {journal} {\bibinfo  {journal} {WIREs Comput
  Mol Sci}\ }\textbf {\bibinfo {volume} {8}} (\bibinfo {year} {2018}),\
  10.1002/wcms.1340}\BibitemShut {NoStop}%
\bibitem [{\citenamefont {Sun}\ \emph {et~al.}(2020)\citenamefont {Sun},
  \citenamefont {Zhang}, \citenamefont {Banerjee}, \citenamefont {Bao},
  \citenamefont {Barbry}, \citenamefont {Blunt}, \citenamefont {Bogdanov},
  \citenamefont {Booth}, \citenamefont {Chen}, \citenamefont {Cui},
  \citenamefont {Eriksen}, \citenamefont {Gao}, \citenamefont {Guo},
  \citenamefont {Hermann}, \citenamefont {Hermes}, \citenamefont {Koh},
  \citenamefont {Koval}, \citenamefont {Lehtola}, \citenamefont {Li},
  \citenamefont {Liu}, \citenamefont {Mardirossian}, \citenamefont {McClain},
  \citenamefont {Motta}, \citenamefont {Mussard}, \citenamefont {Pham},
  \citenamefont {Pulkin}, \citenamefont {Purwanto}, \citenamefont {Robinson},
  \citenamefont {Ronca}, \citenamefont {Sayfutyarova}, \citenamefont
  {Scheurer}, \citenamefont {Schurkus}, \citenamefont {Smith}, \citenamefont
  {Sun}, \citenamefont {Sun}, \citenamefont {Upadhyay}, \citenamefont {Wagner},
  \citenamefont {Wang}, \citenamefont {White}, \citenamefont {Whitfield},
  \citenamefont {Williamson}, \citenamefont {Wouters}, \citenamefont {Yang},
  \citenamefont {Yu}, \citenamefont {Zhu}, \citenamefont {Berkelbach},
  \citenamefont {Sharma}, \citenamefont {Sokolov},\ and\ \citenamefont
  {Chan}}]{sun20}%
  \BibitemOpen
  \bibfield  {author} {\bibinfo {author} {\bibfnamefont {Q.}~\bibnamefont
  {Sun}}, \bibinfo {author} {\bibfnamefont {X.}~\bibnamefont {Zhang}}, \bibinfo
  {author} {\bibfnamefont {S.}~\bibnamefont {Banerjee}}, \bibinfo {author}
  {\bibfnamefont {P.}~\bibnamefont {Bao}}, \bibinfo {author} {\bibfnamefont
  {M.}~\bibnamefont {Barbry}}, \bibinfo {author} {\bibfnamefont {N.~S.}\
  \bibnamefont {Blunt}}, \bibinfo {author} {\bibfnamefont {N.~A.}\ \bibnamefont
  {Bogdanov}}, \bibinfo {author} {\bibfnamefont {G.~H.}\ \bibnamefont {Booth}},
  \bibinfo {author} {\bibfnamefont {J.}~\bibnamefont {Chen}}, \bibinfo {author}
  {\bibfnamefont {Z.-H.}\ \bibnamefont {Cui}}, \bibinfo {author} {\bibfnamefont
  {J.~J.}\ \bibnamefont {Eriksen}}, \bibinfo {author} {\bibfnamefont
  {Y.}~\bibnamefont {Gao}}, \bibinfo {author} {\bibfnamefont {S.}~\bibnamefont
  {Guo}}, \bibinfo {author} {\bibfnamefont {J.}~\bibnamefont {Hermann}},
  \bibinfo {author} {\bibfnamefont {M.~R.}\ \bibnamefont {Hermes}}, \bibinfo
  {author} {\bibfnamefont {K.}~\bibnamefont {Koh}}, \bibinfo {author}
  {\bibfnamefont {P.}~\bibnamefont {Koval}}, \bibinfo {author} {\bibfnamefont
  {S.}~\bibnamefont {Lehtola}}, \bibinfo {author} {\bibfnamefont
  {Z.}~\bibnamefont {Li}}, \bibinfo {author} {\bibfnamefont {J.}~\bibnamefont
  {Liu}}, \bibinfo {author} {\bibfnamefont {N.}~\bibnamefont {Mardirossian}},
  \bibinfo {author} {\bibfnamefont {J.~D.}\ \bibnamefont {McClain}}, \bibinfo
  {author} {\bibfnamefont {M.}~\bibnamefont {Motta}}, \bibinfo {author}
  {\bibfnamefont {B.}~\bibnamefont {Mussard}}, \bibinfo {author} {\bibfnamefont
  {H.~Q.}\ \bibnamefont {Pham}}, \bibinfo {author} {\bibfnamefont
  {A.}~\bibnamefont {Pulkin}}, \bibinfo {author} {\bibfnamefont
  {W.}~\bibnamefont {Purwanto}}, \bibinfo {author} {\bibfnamefont {P.~J.}\
  \bibnamefont {Robinson}}, \bibinfo {author} {\bibfnamefont {E.}~\bibnamefont
  {Ronca}}, \bibinfo {author} {\bibfnamefont {E.~R.}\ \bibnamefont
  {Sayfutyarova}}, \bibinfo {author} {\bibfnamefont {M.}~\bibnamefont
  {Scheurer}}, \bibinfo {author} {\bibfnamefont {H.~F.}\ \bibnamefont
  {Schurkus}}, \bibinfo {author} {\bibfnamefont {J.~E.~T.}\ \bibnamefont
  {Smith}}, \bibinfo {author} {\bibfnamefont {C.}~\bibnamefont {Sun}}, \bibinfo
  {author} {\bibfnamefont {S.-N.}\ \bibnamefont {Sun}}, \bibinfo {author}
  {\bibfnamefont {S.}~\bibnamefont {Upadhyay}}, \bibinfo {author}
  {\bibfnamefont {L.~K.}\ \bibnamefont {Wagner}}, \bibinfo {author}
  {\bibfnamefont {X.}~\bibnamefont {Wang}}, \bibinfo {author} {\bibfnamefont
  {A.}~\bibnamefont {White}}, \bibinfo {author} {\bibfnamefont {J.~D.}\
  \bibnamefont {Whitfield}}, \bibinfo {author} {\bibfnamefont {M.~J.}\
  \bibnamefont {Williamson}}, \bibinfo {author} {\bibfnamefont
  {S.}~\bibnamefont {Wouters}}, \bibinfo {author} {\bibfnamefont
  {J.}~\bibnamefont {Yang}}, \bibinfo {author} {\bibfnamefont {J.~M.}\
  \bibnamefont {Yu}}, \bibinfo {author} {\bibfnamefont {T.}~\bibnamefont
  {Zhu}}, \bibinfo {author} {\bibfnamefont {T.~C.}\ \bibnamefont {Berkelbach}},
  \bibinfo {author} {\bibfnamefont {S.}~\bibnamefont {Sharma}}, \bibinfo
  {author} {\bibfnamefont {A.~Y.}\ \bibnamefont {Sokolov}}, \ and\ \bibinfo
  {author} {\bibfnamefont {G.~K.-L.}\ \bibnamefont {Chan}},\ }\href {\doibase
  10.1063/5.0006074} {\bibfield  {journal} {\bibinfo  {journal} {J. Chem.
  Phys.}\ }\textbf {\bibinfo {volume} {153}},\ \bibinfo {pages} {024109}
  (\bibinfo {year} {2020})}\BibitemShut {NoStop}%
\bibitem [{\citenamefont {Ked\v{z}uch}, \citenamefont {Milko},\ and\
  \citenamefont {Noga}(2005)}]{kedzuch05}%
  \BibitemOpen
  \bibfield  {author} {\bibinfo {author} {\bibfnamefont {S.}~\bibnamefont
  {Ked\v{z}uch}}, \bibinfo {author} {\bibfnamefont {M.}~\bibnamefont {Milko}},
  \ and\ \bibinfo {author} {\bibfnamefont {J.}~\bibnamefont {Noga}},\ }\href
  {\doibase 10.1002/qua.20744} {\bibfield  {journal} {\bibinfo  {journal} {Int.
  J. Quantum Chem.}\ }\textbf {\bibinfo {volume} {105}},\ \bibinfo {pages}
  {929} (\bibinfo {year} {2005})}\BibitemShut {NoStop}%
\bibitem [{\citenamefont {Werner}, \citenamefont {Adler},\ and\ \citenamefont
  {Manby}(2007)}]{werner07}%
  \BibitemOpen
  \bibfield  {author} {\bibinfo {author} {\bibfnamefont {H.-J.}\ \bibnamefont
  {Werner}}, \bibinfo {author} {\bibfnamefont {T.~B.}\ \bibnamefont {Adler}}, \
  and\ \bibinfo {author} {\bibfnamefont {F.~R.}\ \bibnamefont {Manby}},\ }\href
  {\doibase 10.1063/1.2712434} {\bibfield  {journal} {\bibinfo  {journal} {J.
  Chem. Phys.}\ }\textbf {\bibinfo {volume} {126}},\ \bibinfo {pages} {164102}
  (\bibinfo {year} {2007})}\BibitemShut {NoStop}%
\bibitem [{\citenamefont {Bomble}, \citenamefont {Vázquez},\ and\
  \citenamefont {Kállay}(2006)}]{bomble06}%
  \BibitemOpen
  \bibfield  {author} {\bibinfo {author} {\bibfnamefont {Y.~J.}\ \bibnamefont
  {Bomble}}, \bibinfo {author} {\bibfnamefont {J.}~\bibnamefont {Vázquez}}, \
  and\ \bibinfo {author} {\bibfnamefont {M.}~\bibnamefont {Kállay}},\
  }\href@noop {} {\bibfield  {journal} {\bibinfo  {journal} {J. Chem. Phys.}\
  }\textbf {\bibinfo {volume} {125}},\ \bibinfo {pages} {064108} (\bibinfo
  {year} {2006})}\BibitemShut {NoStop}%
\bibitem [{\citenamefont {Harding}\ \emph {et~al.}(2008)\citenamefont
  {Harding}, \citenamefont {Vázquez}, \citenamefont {Ruscic}, \citenamefont
  {Wilson}, \citenamefont {Gauss},\ and\ \citenamefont {Stanton}}]{harding08}%
  \BibitemOpen
  \bibfield  {author} {\bibinfo {author} {\bibfnamefont {M.~E.}\ \bibnamefont
  {Harding}}, \bibinfo {author} {\bibfnamefont {J.}~\bibnamefont {Vázquez}},
  \bibinfo {author} {\bibfnamefont {B.}~\bibnamefont {Ruscic}}, \bibinfo
  {author} {\bibfnamefont {A.~K.}\ \bibnamefont {Wilson}}, \bibinfo {author}
  {\bibfnamefont {J.}~\bibnamefont {Gauss}}, \ and\ \bibinfo {author}
  {\bibfnamefont {J.~F.}\ \bibnamefont {Stanton}},\ }\href {\doibase
  10.1063/1.2835612} {\bibfield  {journal} {\bibinfo  {journal} {J. Chem.
  Phys.}\ }\textbf {\bibinfo {volume} {128}},\ \bibinfo {pages} {114111}
  (\bibinfo {year} {2008})}\BibitemShut {NoStop}%
\bibitem [{\citenamefont {Thorpe}\ \emph {et~al.}(2019)\citenamefont {Thorpe},
  \citenamefont {Lopez}, \citenamefont {Nguyen}, \citenamefont {Baraban},
  \citenamefont {Bross}, \citenamefont {Ruscic},\ and\ \citenamefont
  {Stanton}}]{thorpe19}%
  \BibitemOpen
  \bibfield  {author} {\bibinfo {author} {\bibfnamefont {J.~H.}\ \bibnamefont
  {Thorpe}}, \bibinfo {author} {\bibfnamefont {C.~A.}\ \bibnamefont {Lopez}},
  \bibinfo {author} {\bibfnamefont {T.~L.}\ \bibnamefont {Nguyen}}, \bibinfo
  {author} {\bibfnamefont {J.~H.}\ \bibnamefont {Baraban}}, \bibinfo {author}
  {\bibfnamefont {D.~H.}\ \bibnamefont {Bross}}, \bibinfo {author}
  {\bibfnamefont {B.}~\bibnamefont {Ruscic}}, \ and\ \bibinfo {author}
  {\bibfnamefont {J.~F.}\ \bibnamefont {Stanton}},\ }\href {\doibase
  10.1063/1.5095937} {\bibfield  {journal} {\bibinfo  {journal} {J. Chem.
  Phys.}\ }\textbf {\bibinfo {volume} {150}},\ \bibinfo {pages} {224102}
  (\bibinfo {year} {2019})}\BibitemShut {NoStop}%
\bibitem [{\citenamefont {Feller}(1992)}]{feller92}%
  \BibitemOpen
  \bibfield  {author} {\bibinfo {author} {\bibfnamefont {D.}~\bibnamefont
  {Feller}},\ }\href {\doibase 10.1063/1.462652} {\bibfield  {journal}
  {\bibinfo  {journal} {J. Chem. Phys.}\ }\textbf {\bibinfo {volume} {96}},\
  \bibinfo {pages} {6104} (\bibinfo {year} {1992})}\BibitemShut {NoStop}%
\bibitem [{\citenamefont {Feller}(1993)}]{feller93}%
  \BibitemOpen
  \bibfield  {author} {\bibinfo {author} {\bibfnamefont {D.}~\bibnamefont
  {Feller}},\ }\href {\doibase 10.1063/1.464749} {\bibfield  {journal}
  {\bibinfo  {journal} {J. Chem. Phys.}\ }\textbf {\bibinfo {volume} {98}},\
  \bibinfo {pages} {7059} (\bibinfo {year} {1993})}\BibitemShut {NoStop}%
\bibitem [{\citenamefont {Helgaker}\ \emph {et~al.}(1997)\citenamefont
  {Helgaker}, \citenamefont {Klopper}, \citenamefont {Koch},\ and\
  \citenamefont {Noga}}]{helgaker97}%
  \BibitemOpen
  \bibfield  {author} {\bibinfo {author} {\bibfnamefont {T.}~\bibnamefont
  {Helgaker}}, \bibinfo {author} {\bibfnamefont {W.}~\bibnamefont {Klopper}},
  \bibinfo {author} {\bibfnamefont {H.}~\bibnamefont {Koch}}, \ and\ \bibinfo
  {author} {\bibfnamefont {J.}~\bibnamefont {Noga}},\ }\href {\doibase
  10.1063/1.473863} {\bibfield  {journal} {\bibinfo  {journal} {J. Chem.
  Phys.}\ }\textbf {\bibinfo {volume} {106}},\ \bibinfo {pages} {9639}
  (\bibinfo {year} {1997})}\BibitemShut {NoStop}%
\bibitem [{\citenamefont {Werner}\ and\ \citenamefont {Knowles}(1988)}]{WK88}%
  \BibitemOpen
  \bibfield  {author} {\bibinfo {author} {\bibfnamefont {H.-J.}\ \bibnamefont
  {Werner}}\ and\ \bibinfo {author} {\bibfnamefont {P.~J.}\ \bibnamefont
  {Knowles}},\ }\href {\doibase 10.1063/1.455556} {\bibfield  {journal}
  {\bibinfo  {journal} {J. Chem. Phys.}\ }\textbf {\bibinfo {volume} {89}},\
  \bibinfo {pages} {5803} (\bibinfo {year} {1988})}\BibitemShut {NoStop}%
\bibitem [{\citenamefont {Knowles}\ and\ \citenamefont {Werner}(1988)}]{KW88}%
  \BibitemOpen
  \bibfield  {author} {\bibinfo {author} {\bibfnamefont {P.~J.}\ \bibnamefont
  {Knowles}}\ and\ \bibinfo {author} {\bibfnamefont {H.-J.}\ \bibnamefont
  {Werner}},\ }\href {\doibase 10.1016/0009-2614(88)87412-8} {\bibfield
  {journal} {\bibinfo  {journal} {Chem. Phys. Letters}\ }\textbf {\bibinfo
  {volume} {145}},\ \bibinfo {pages} {514} (\bibinfo {year}
  {1988})}\BibitemShut {NoStop}%
\bibitem [{\citenamefont {Shiozaki}, \citenamefont {Knizia},\ and\
  \citenamefont {Werner}(2011)}]{Shiozaki:2011}%
  \BibitemOpen
  \bibfield  {author} {\bibinfo {author} {\bibfnamefont {T.}~\bibnamefont
  {Shiozaki}}, \bibinfo {author} {\bibfnamefont {G.}~\bibnamefont {Knizia}}, \
  and\ \bibinfo {author} {\bibfnamefont {H.-J.}\ \bibnamefont {Werner}},\
  }\href@noop {} {\bibfield  {journal} {\bibinfo  {journal} {J. Chem. Phys.}\
  }\textbf {\bibinfo {volume} {134}},\ \bibinfo {pages} {034113} (\bibinfo
  {year} {2011})}\BibitemShut {NoStop}%
\bibitem [{\citenamefont {Shiozaki}\ and\ \citenamefont
  {Werner}(2011)}]{Shiozaki:2011a}%
  \BibitemOpen
  \bibfield  {author} {\bibinfo {author} {\bibfnamefont {T.}~\bibnamefont
  {Shiozaki}}\ and\ \bibinfo {author} {\bibfnamefont {H.-J.}\ \bibnamefont
  {Werner}},\ }\href@noop {} {\bibfield  {journal} {\bibinfo  {journal} {J.
  Chem. Phys.}\ }\textbf {\bibinfo {volume} {134}},\ \bibinfo {pages} {184104}
  (\bibinfo {year} {2011})}\BibitemShut {NoStop}%
\bibitem [{\citenamefont {Tew}\ and\ \citenamefont {Klopper}(2005)}]{tew05}%
  \BibitemOpen
  \bibfield  {author} {\bibinfo {author} {\bibfnamefont {D.~P.}\ \bibnamefont
  {Tew}}\ and\ \bibinfo {author} {\bibfnamefont {W.}~\bibnamefont {Klopper}},\
  }\href {\doibase 10.1063/1.1999632} {\bibfield  {journal} {\bibinfo
  {journal} {J. Chem. Phys.}\ }\textbf {\bibinfo {volume} {123}},\ \bibinfo
  {pages} {074101} (\bibinfo {year} {2005})}\BibitemShut {NoStop}%
\end{thebibliography}%

\end{document}